\documentclass[%
 reprint,
 amsmath,amssymb,
 aps,
]{revtex4-1}

\usepackage[utf8]{inputenc}
\usepackage{graphicx}
\usepackage{dcolumn}
\usepackage{bm}
\usepackage{float}
\usepackage{amsmath}
\usepackage{graphicx,subfigure,epsfig}
\usepackage{hyperref}
\hypersetup{
    colorlinks=true,
    citecolor=blue,
    linkcolor=blue,
    filecolor=magenta,
    urlcolor=cyan,}

\newcommand{\be}{\begin{equation}}
\newcommand{\ee}{\end{equation}}

\begin{document}

\preprint{APS/123-QED}

\title{Ringing of a black hole in a dark matter halo}
\author{Dong Liu}
 \email{gs.dongliu19@gzu.edu.cn}
\author{Yi Yang}
\email{gs.yangyi17@gzu.edu.cn}
\author{Shurui Wu}%
\author{Yujia Xing}%
\author{Zhaoyi Xu}%
 \email{zyxu@gzu.edu.cn (Corresponding author)}
\author{Zheng-Wen Long}%
 \email{zwlong@gzu.edu.cn (Corresponding author)}

\affiliation{%
College of Physics, Guizhou University, Guiyang, 550025, China}%

\date{\today}

\begin{abstract}
Recently, we obtained the simple metrics of a spherically symmetric black hole in a dark matter halo, and extended to the case of rotation.  As the characteristic sound of black holes, quasinormal modes (QNMs) are one of the important means to understand black holes currently. Based on these two metrics of a spherically symmetric black hole, we study the QNMs of cold dark matter (CDM) and scalar field dark matter (SFDM) models using the methods of the material field perturbations and the gravitational perturbation, and make comparisons with the Schwarzschild black hole. Our results show that black hole QNMs of CDM and SFDM in a dark matter halo are different from the Schwarzschild black hole, unlike a Schwarzschild black hole with a prominent power-law tail. The different kinds models of dark matter can be distinguished by their QNMs. The time of QNMs ringing and frequencies increase with increasing parameter $l$. The overall QNMs of CDM are stronger than that of SFDM in the same condition, which is easier to be detected. In addition, QNMs frequencies using the sixth-order WKB method and the Prony method are in good agreement.
\end{abstract}

\maketitle


\section{Introduction}\label{sec:1}
Astronomical observations show that our Universe mainly consists of three kinds of essential matters: 4.9\% baryonic matter, 26.8\% dark matter, and 68.3\% dark energy \cite{P . A. R.}. For heavenly bodies in the Universe, dark matter's effects are extremely important. The most mainstream dark matter model presently is the cold dark matter model \cite{J. F. Navarr,J. F. Navarro}, but it has a series of observation difficulties in the small-scale structure of the Universe \cite{Friedmann D E,Robles V H}. So various dark matter models for studying dark matter  have been proposed to study dark matter, such as a warm dark matter (WDM) model \cite{P . Bode,P. Coln}, a Bose-Einstein condensation (BEC) model \cite{W. H.,S.-J.}, and a self-interacting dark matter (SIDM) model \cite{D. N. Spergel,M. Kaplinghat}. Among these models, astronomers are most interested in the distribution of dark matter corresponding to the dark matter model. The distribution of dark matter in the large-scale structure of the galaxy is clear presently \cite{P . A. R.}, but it is often unclear near the nucleus of the galaxy and in the supermassive black holes or intermediate massive black holes. Therefore, researching the distribution of dark matter near black holes will be an interesting and important question. Generally speaking, for a supermassive black hole in the Universe, its strong gravitation can cause the density of dark matter near the black hole to increase sharply, creating a ``spike" phenomenon \cite{P. Gondolo,L. Sadeghian,B.D. Fields}.\\
\indent On the other hand, a black hole(BH) is celestial body predicted by general relativity. Wheeler announced that an isolated black hole can be described by its mass, angular momentum, and charge\cite{mc}. However, there are almost no isolated black holes in the real Universe. There may be various complicated matter fields in the black hole. This indicates that the black hole always interacts with the external source field, causing the real black hole to be in a perturbed state. Here, we introduce quasinormal modes (QNMs) of black holes. QNMs are produced when black holes are perturbed\cite{OA}. It is well known that there are three stages of black hole perturbations. The first stage is the initial stage, and the second is the QNMs ringing, which hides the QNMs frequencies. The third is the power-law tail. QNMs can be obtained by solving black hole perturbation equations. The solution of the equation is represented by the pure outgoing wave at infinity and the pure ingoing wave at the event horizon\cite{s.f,i.s,d.d,c.c,p.a}. As the characteristic sound of black holes, QNMs are one of the important means to understand black holes currently\cite{R. A. Konoplya*}. For research on QNMs, papers can be referenced\cite{h.p,e.b,b.t,s.a,m.m}. From black hole QNMs, we can extract frequencies of QNMs\cite{B,S,S.,R}. QNM frequencies are not only related to the hairs of the black hole (mass, charge and angular momentum) but also may identify the existence of the black hole\cite{sm}. QNMs can provide a method to identify black holes in the Universe because they carry characteristic information of black holes\cite{CS,RT}. The QNMs of black holes change with parameters changed. Therefore, we can use QNMs as a tool to explore and analyze some inherent properties of a black hole. Now, the detections of a gravitational wave \cite{1,2,4,5,6} have opened up new doors of opportunity for physics research. In the future plan of a gravitational wave, QNMs may be detected, and the effects of dark matter on a black hole will also be reflected as QNMs are discovered. It can be seen that using QNMs to study the distribution of dark matter near black holes is a very important topic, which may help us better solve the distribution problem of dark matter near a black hole.\\
\indent In this paper, the emphasis of our work is on studying the effects of QNMs on the distribution of dark matter near black holes. It is well worth noting Refs. \cite{Xu,Xu1,Xu2,Xu3,Xu4}. Through theoretical derivation, we obtained black hole metrics of the static spherically symmetric in a dark matter halo. In these metrics, QNMs will occur in the perturbations of a black hole space-time. We utilize the QNMs to study the characteristics of the dark matter halo near the black hole, and make comparisons with the Schwarzschild black hole(SCHW). In addition, QNMs can help us distinguish the geometries of different dark matter models \cite{DUTTAROY}.\\
\indent The paper is organized as follows. In Sec.\ref{sec:2}, we introduce the black hole metrics in a dark matter halo, the equations of motion under the material field perturbations and gravitational perturbation, the WKB method and the finite difference method we used. In Sec.\ref{sec:3}, we present the quasinormal modes under scalar, electromagnetic fields, gravitational perturbation and QNM frequency tables. Finally, Sec.\ref{sec:4} has discussions and conclusions. In this paper we use mostly the units ($G=c=1$).

\section{The methods}\label{sec:2}
\subsection{Material field perturbations of a BH in a dark matter halo}
\label{startsample}
The static spherically symmetric metric of a black hole is usually given in the following form,
\begin{eqnarray}
ds^{2}=-f(r)dt^{2}+\frac{1}{f(r)}dr^{2}+r^{2}(d\theta ^{2}+\sin^{2}\theta d\phi ^{2}).
\label{equ1}
\end{eqnarray}
\indent We consider the metric of the static spherically symmetric black hole under the dark matter halo in Ref. \cite{Xu}:
\indent For cold dark matter(CDM),
\begin{eqnarray}  
f(r)=(1+\frac{r}{R_{\rm c}})^{-\frac{8\pi \rho_{\rm c}R_{\rm c}^{3}}{r}}-\frac{2M}{r},
\label{equ2}
\end{eqnarray}  
\indent For scalar field dark matter(SFDM),
\begin{eqnarray}  
f(r)=exp(-\frac{8\rho _{\rm s}R_{\rm s}^{2}}{\pi }\frac{\sin(\pi r/R_{\rm s})}{\pi r/R_{\rm s}})-\frac{2M}{r},
\label{equ3}
\end{eqnarray}  
where $M$ is the mass of a black hole; $\rho_{\rm c}$ and $\rho_{\rm s}$ are the density of a cosmic period when the halo collapsed; $R_{\rm c}$ and $R_{\rm s}$ are the characteristic radius.

From these two metrics, the difference between them and the Schwarzschild black hole is that there is a transcendental item in $f(r)$. When their transcendental items are equal to $1$, they will be Schwarzschild black holes. Besides, the metrics we used are statically spherically symmetric, and the components of the metrics are unrelated to time, which means that all the $t$ are equal to a constant value are the same \cite{DUTTAROY}.

\indent The motion equation of a massless scalar field is generally a covariant K-G equation \cite{Landau},
\begin{equation}
\frac{1}{\sqrt{-g}}{\partial \mu}(\sqrt{-g}g^{\mu\nu} {\partial \nu}\Phi) =0,     
\label{equ4}               
\end{equation}
and for the motion equation of an electromagnetic field, it has generally the form \cite{R. A. Konoplya*}
\begin{equation} 
\frac{1}{\sqrt{-g}}{\partial \nu }(F_{\rho\sigma}g^{\rho\mu}g^{\sigma\nu}\sqrt{-g})=0,       
\label{equ5}                
\end{equation}
where $ F_{\rho\sigma}={\partial \rho}A^\sigma-{\partial \sigma}A^\rho $, $A_\nu$ is an electromagnetic four-potential.

\begin{figure*}[t!]
\centering
{ 
\label{fig:b}     
\includegraphics[width=0.6\columnwidth]{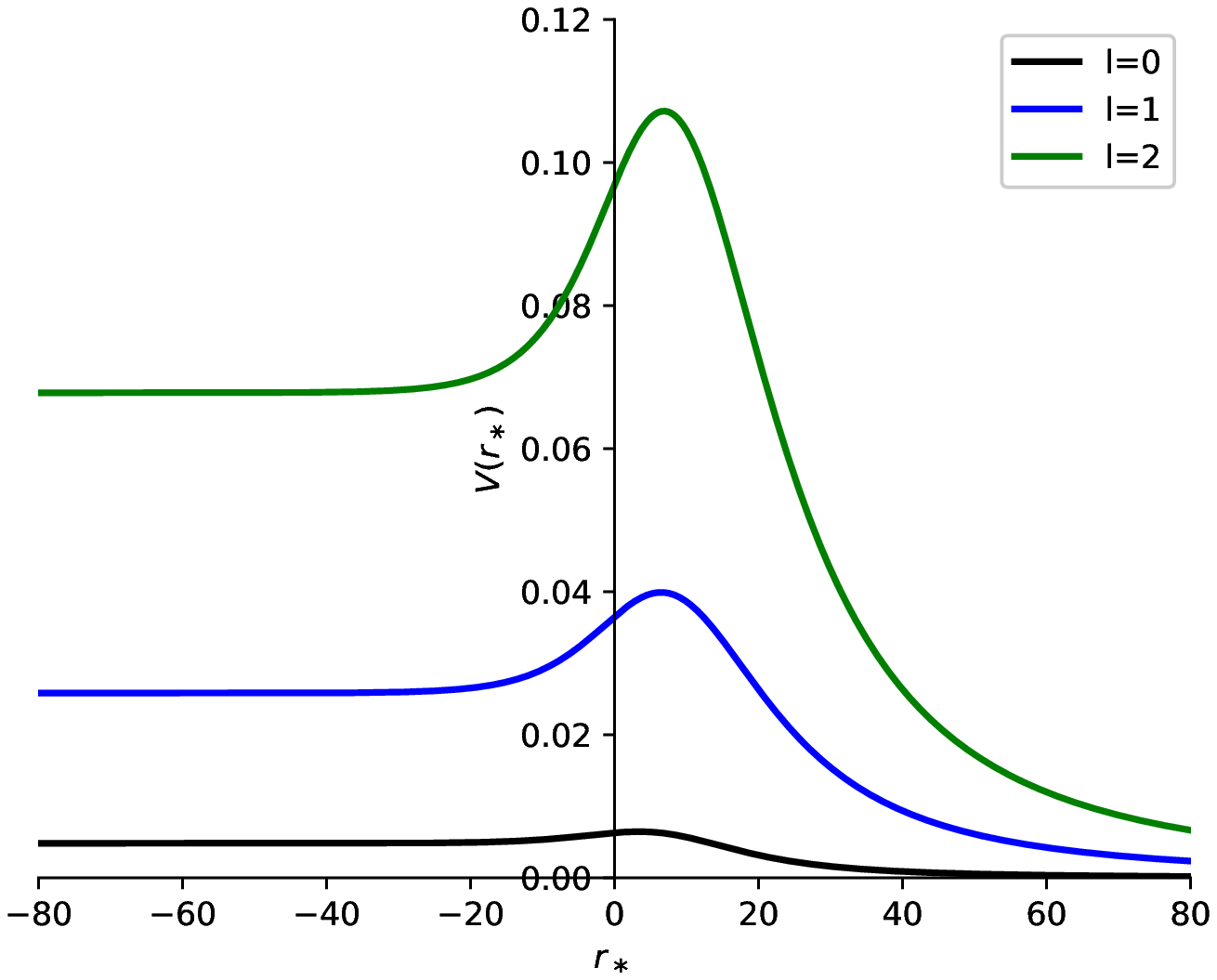}     
} 
{ 
\label{fig:b}     
\includegraphics[width=0.6\columnwidth]{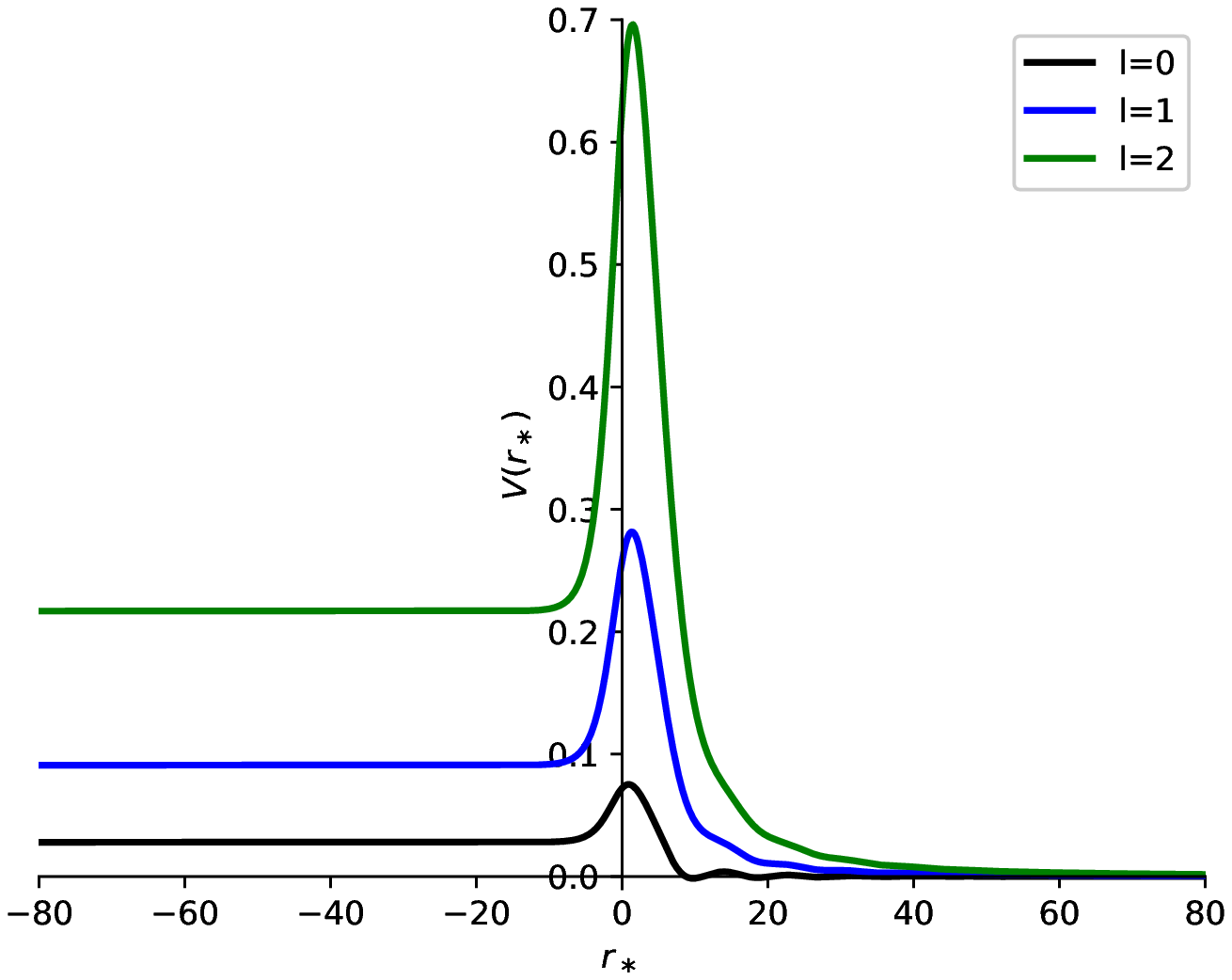}     
}
{ 
\label{fig:b}     
\includegraphics[width=0.6\columnwidth]{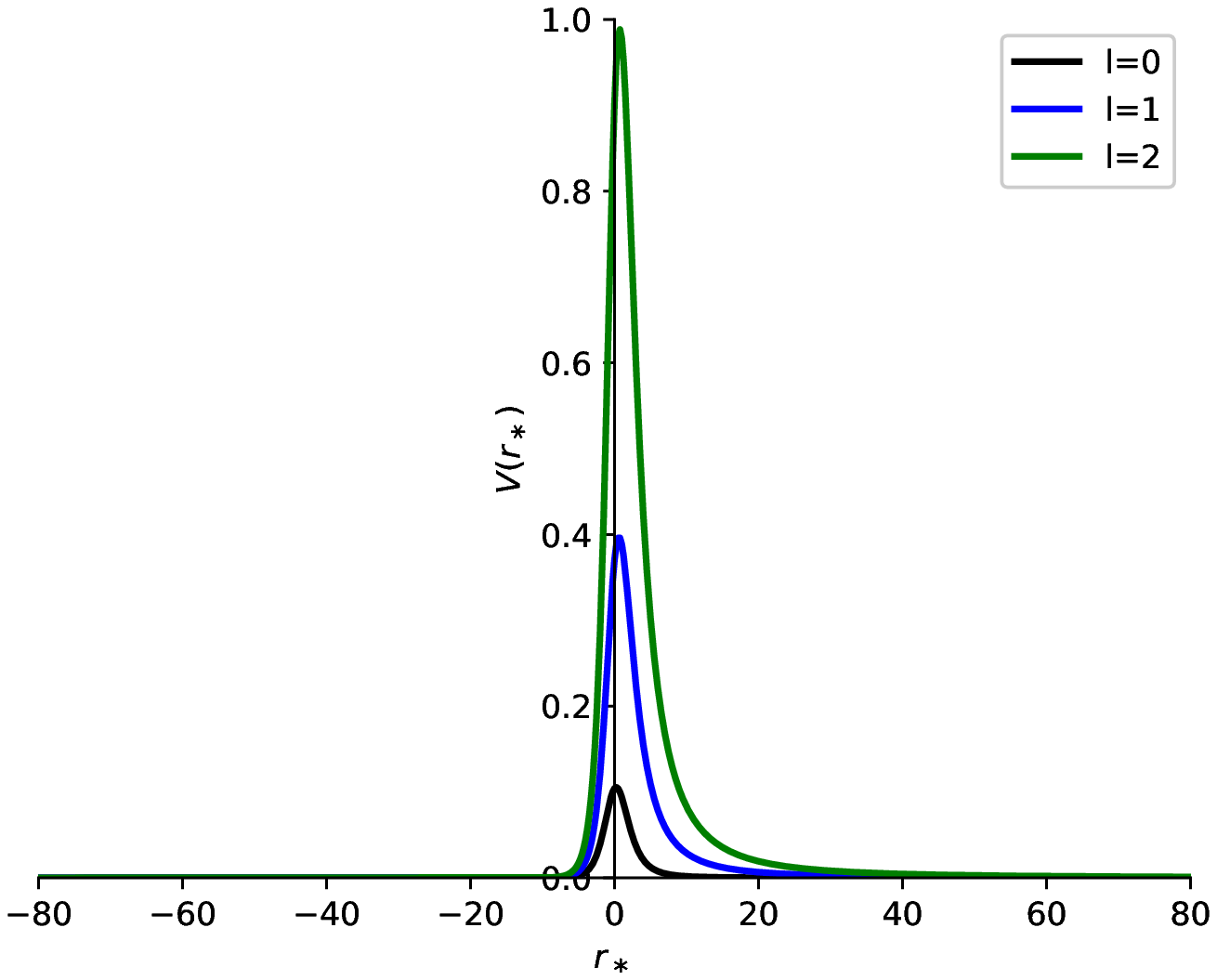}     
}
\caption{The effective potentials of the scalar field with the different $l$. The three panels, from left to right, are CDM, SFDM, SCHW. The parameters we used are $M=0.5, R_{\rm c}=6, \rho_{\rm c}=0.001, R_{\rm s}=3, \rho_{\rm s}=0.01$.}     
\label{fig:1}
\end{figure*}
\begin{figure*}[t!]
\centering
{ 
\label{fig:b}     
\includegraphics[width=0.6\columnwidth]{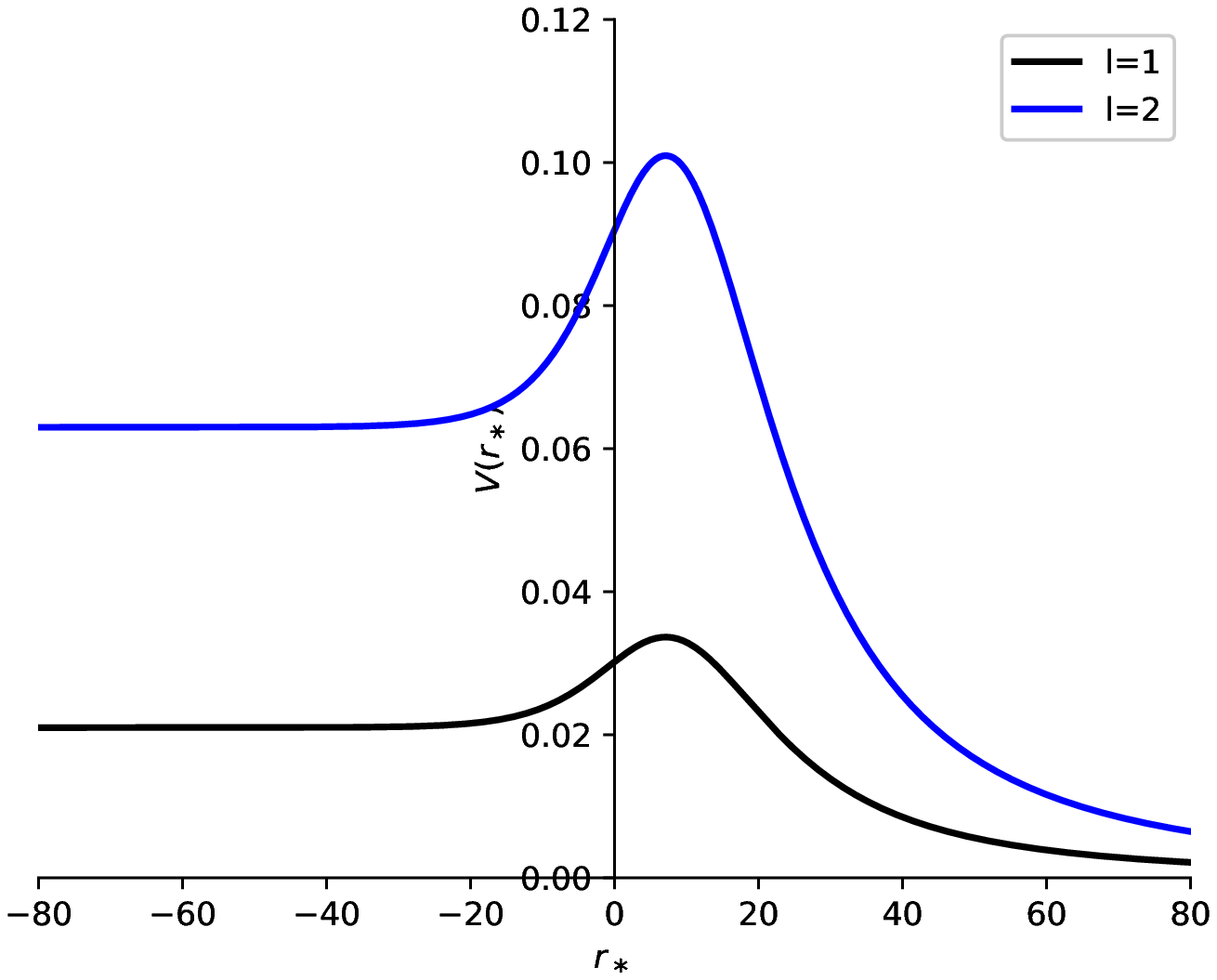}     
} 
{ 
\label{fig:b}     
\includegraphics[width=0.6\columnwidth]{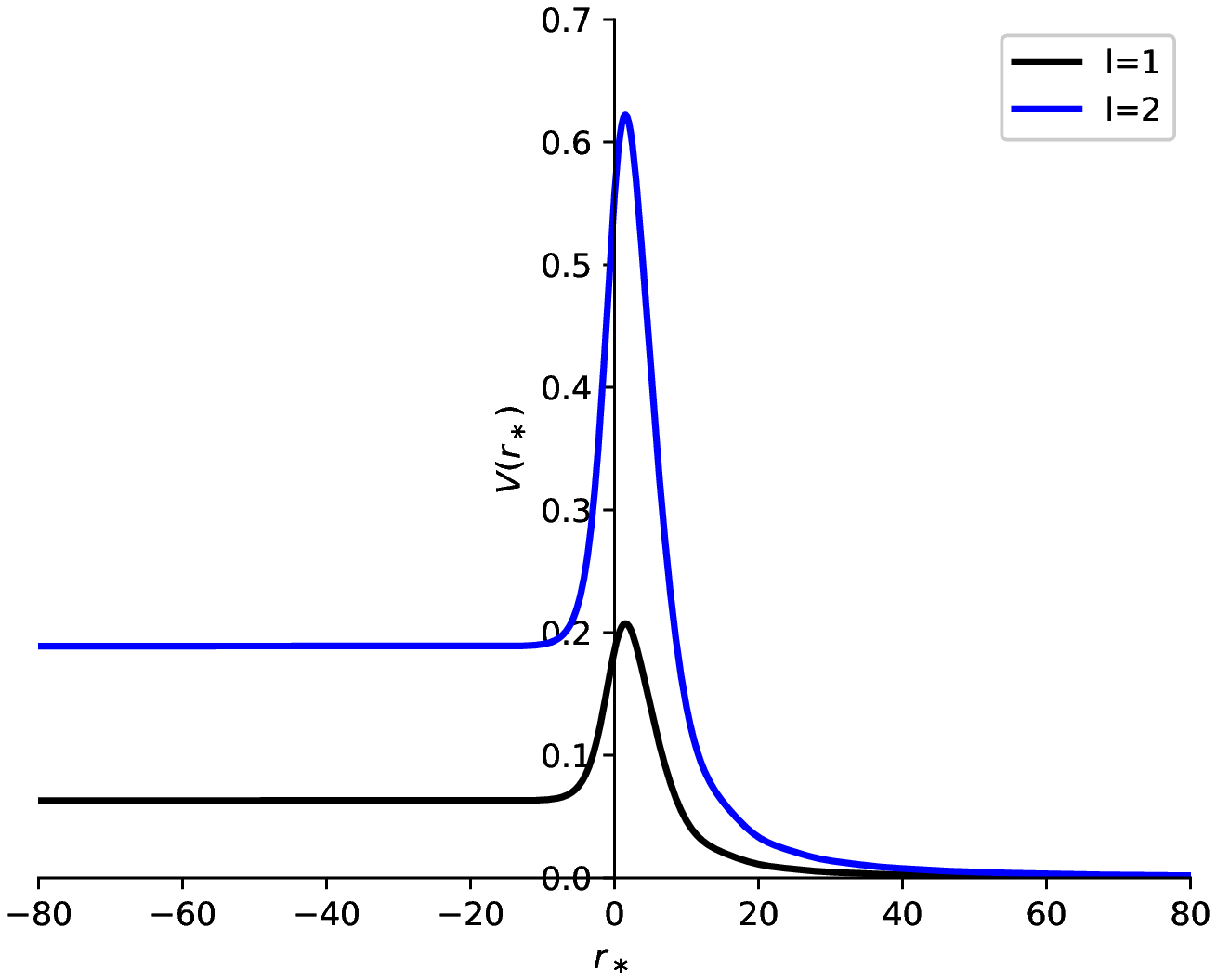}     
}
{ 
\label{fig:b}     
\includegraphics[width=0.6\columnwidth]{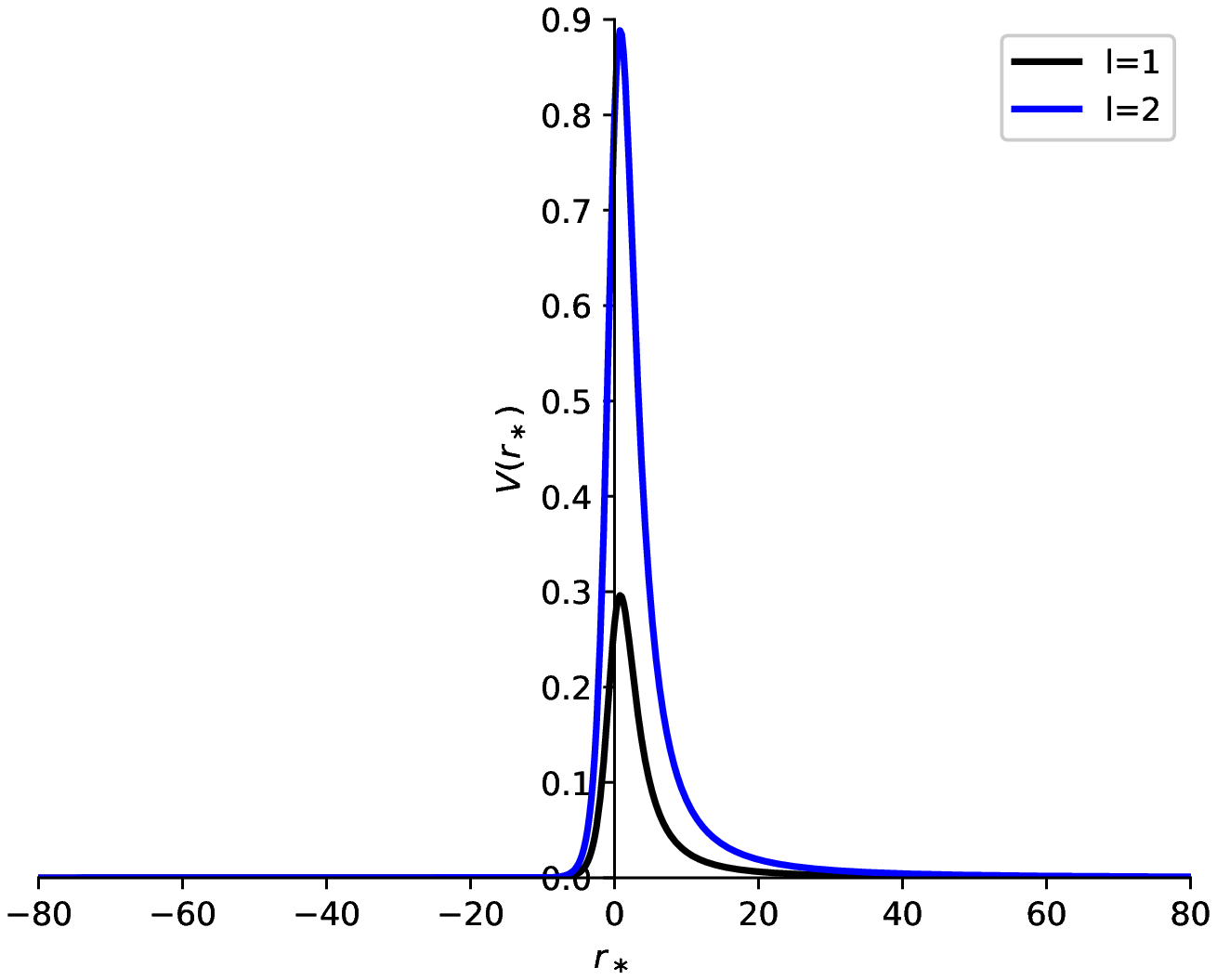}     
}
\caption{The effective potentials of the electromagnetic field with the different $l$. The three panels, from left to right, are CDM, SFDM, SCHW. The parameters we used are $M=0.5, R_{\rm c}=6, \rho_{\rm c}=0.001, R_{\rm s}=3, \rho_{\rm s}=0.01$.}     
\label{fig:2}     
\end{figure*}
\begin{figure*}[t!]
\centering
{  
\includegraphics[width=0.6\columnwidth]{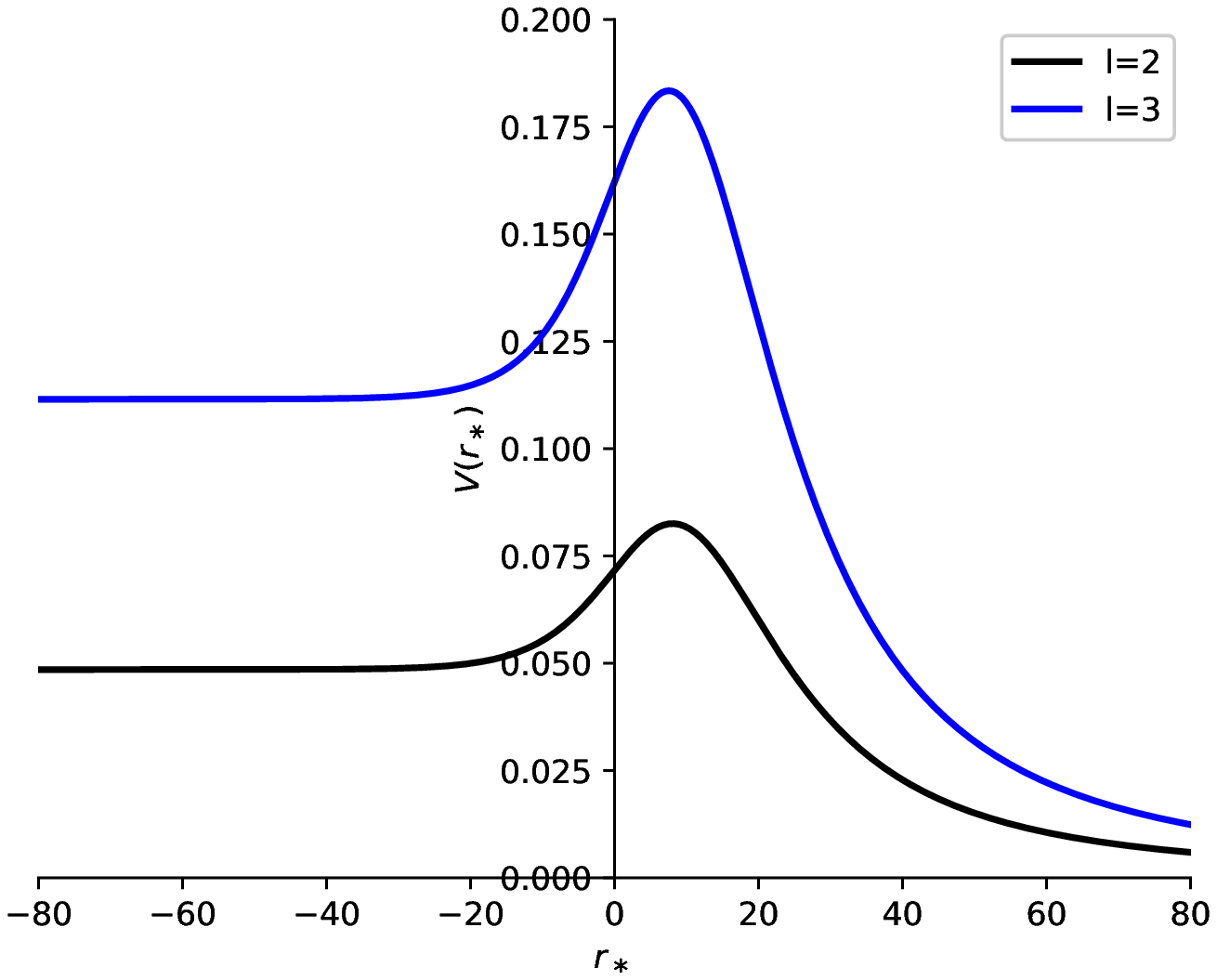}     
} 
{   
\includegraphics[width=0.6\columnwidth]{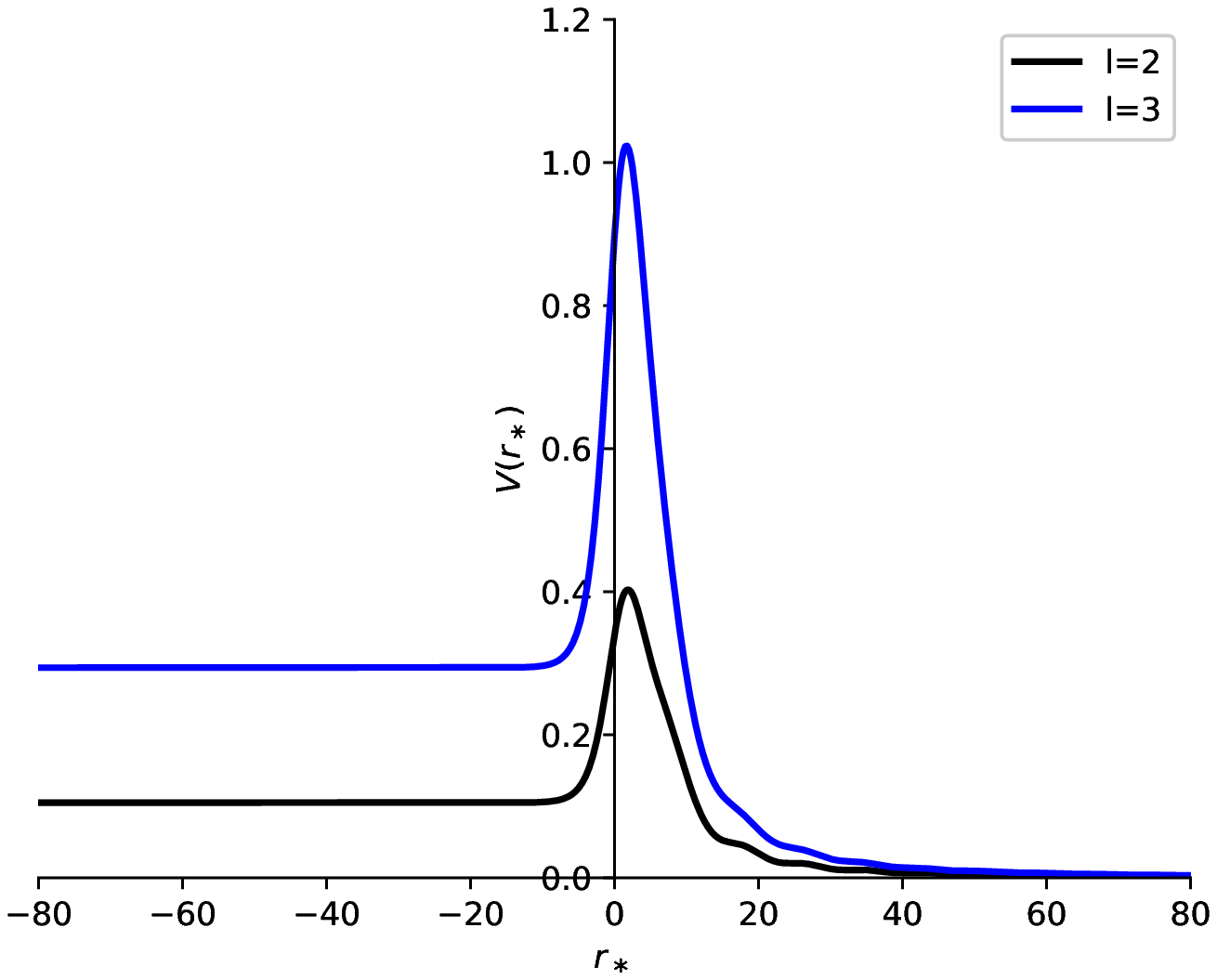}     
}
{    
\includegraphics[width=0.6\columnwidth]{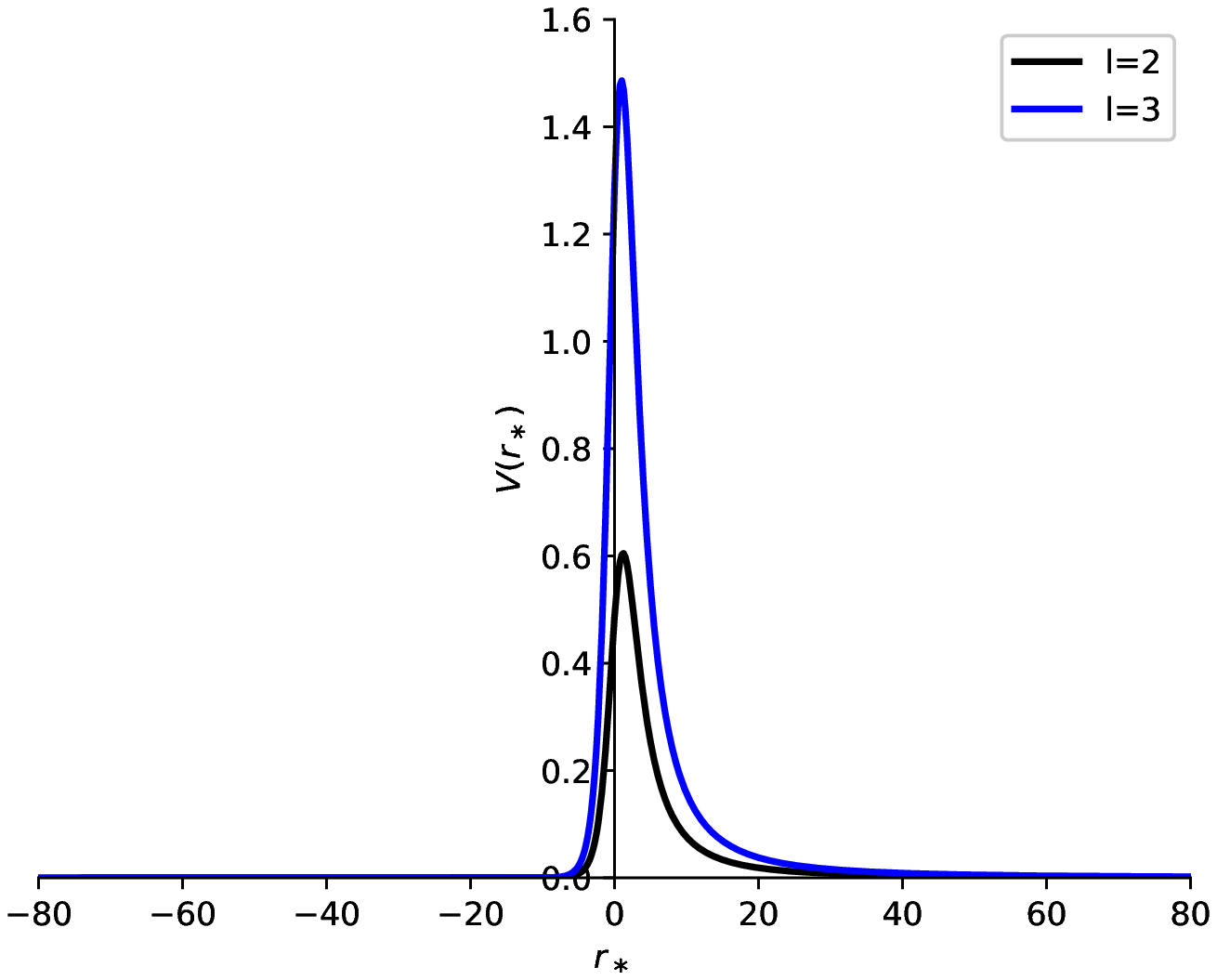}     
}
\caption{The effective potentials of the gravitational field with the different $l$. The three panels, from left to right, are CDM, SFDM, SCHW. The parameter we used is $M=0.5$.}   
\label{fig:3}  
\end{figure*}
\begin{figure*}[t!]
\centering
{    
\includegraphics[width=0.6\columnwidth]{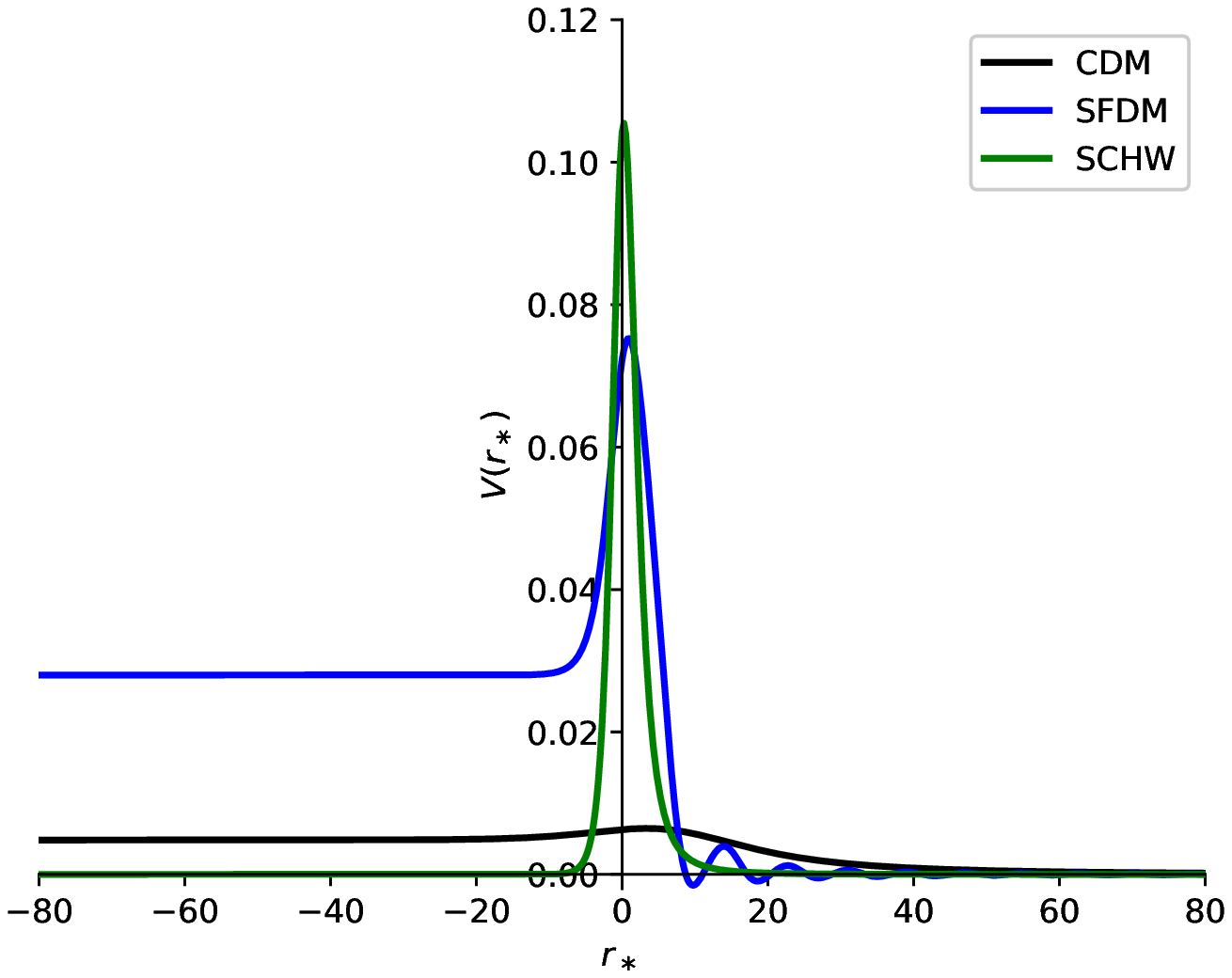}  
}     
{                                                      
\includegraphics[width=0.6\columnwidth]{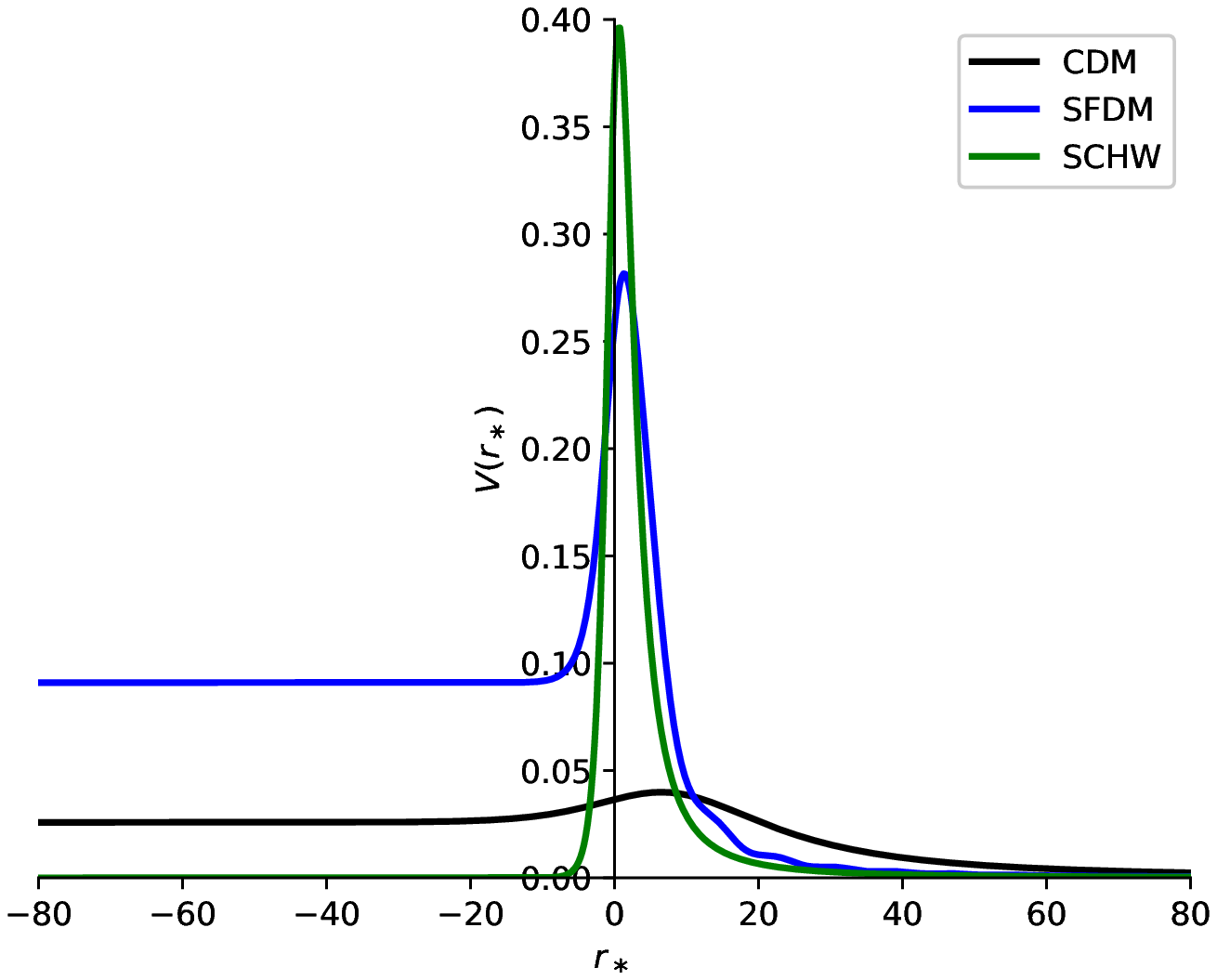}     
}    
{    
\includegraphics[width=0.6\columnwidth]{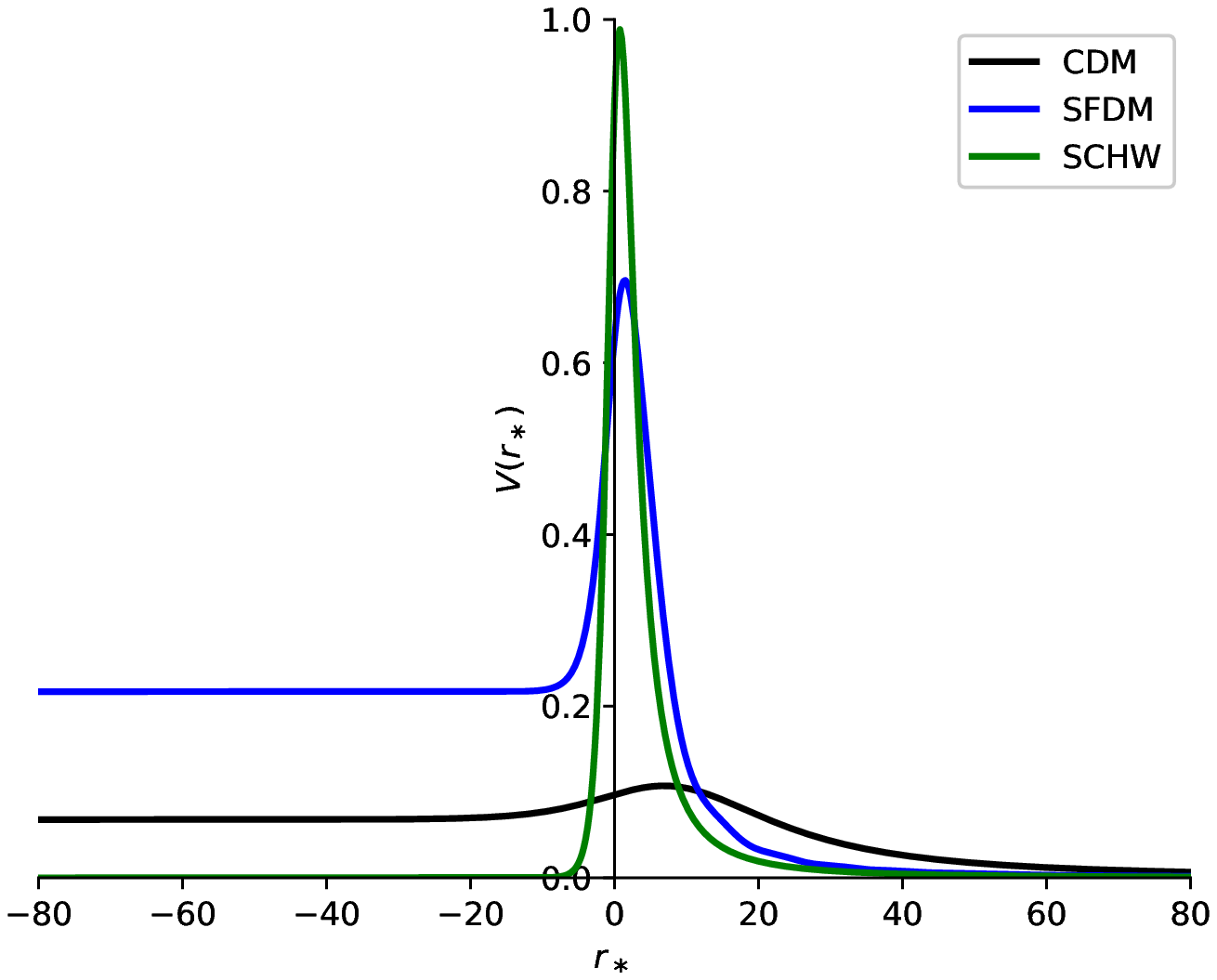}     
}
\caption{The effective potentials of scalar field with the different space-times. The three panels, from left to right, are $l=0, 1, 2$.}     
\label{fig:4}     
\end{figure*}
\begin{figure*}
\centering
{ 
\label{fig:b}     
\includegraphics[width=0.8\columnwidth]{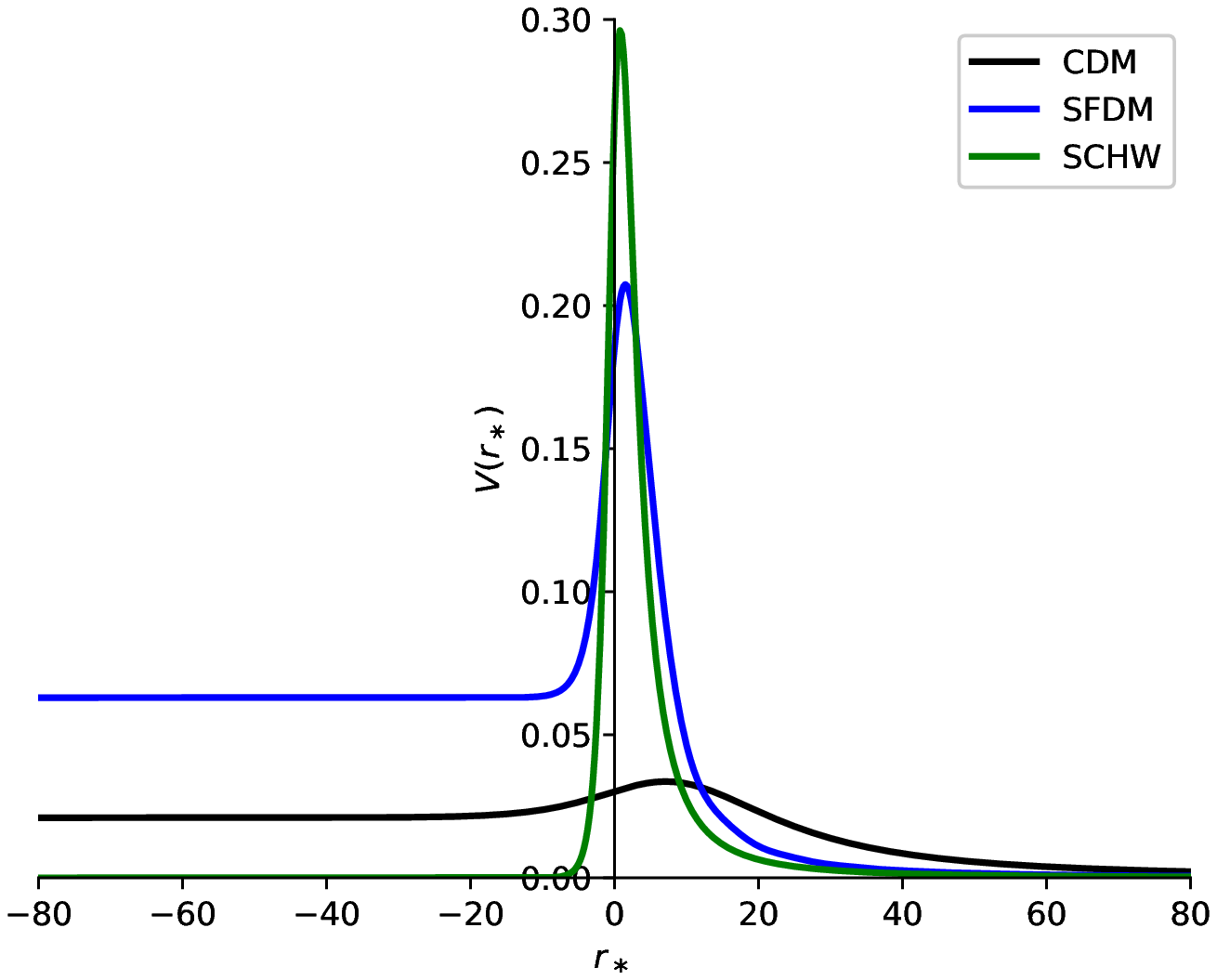}     
} 
{ 
\label{fig:b}     
\includegraphics[width=0.8\columnwidth]{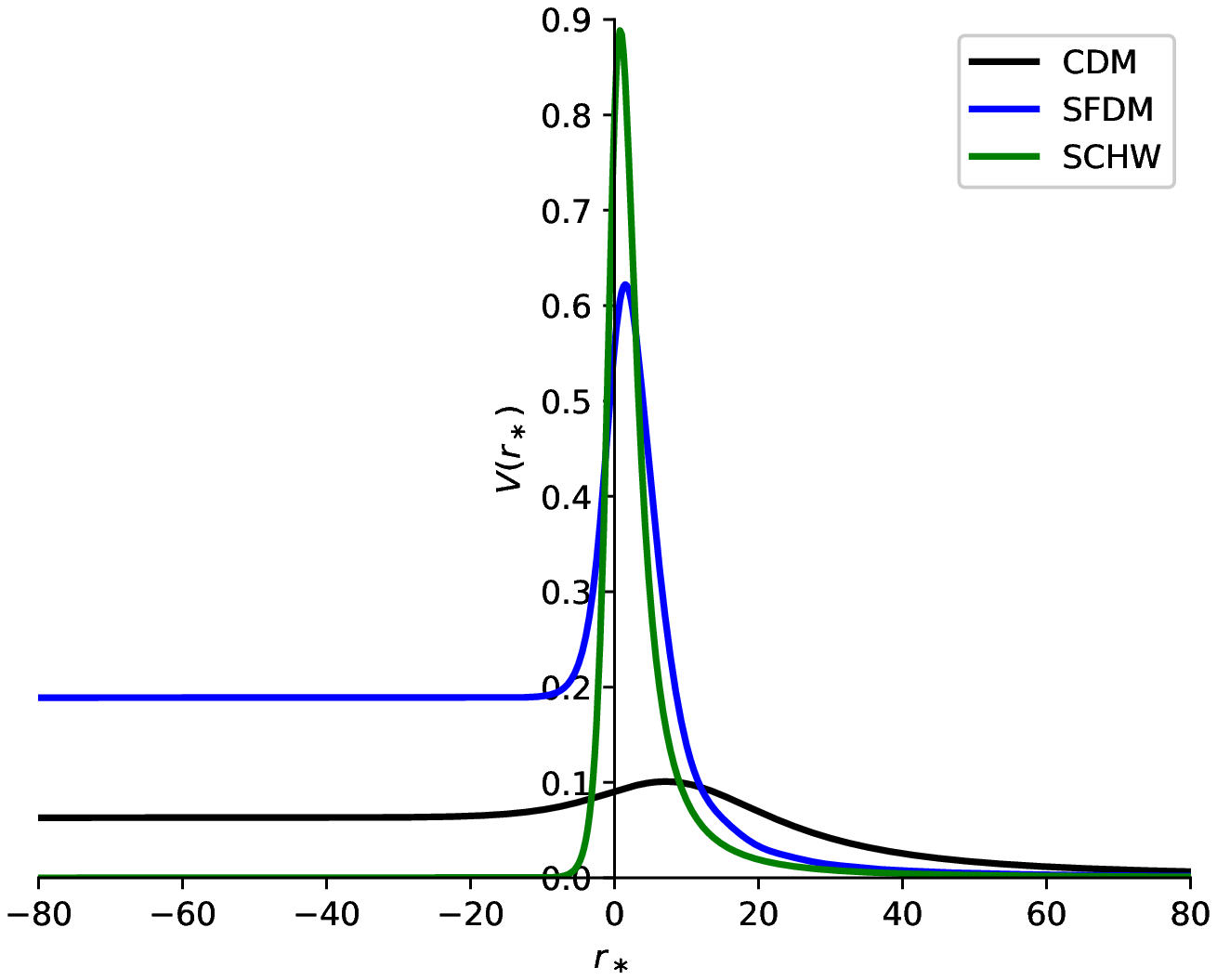}     
}
\caption{The effective potentials of electromagnetic field with the different space-times. The two panels, from left to right, are $l=1, 2$.}     
\label{fig:5}     
\end{figure*}
\begin{figure*}
\centering
{ 
\label{fig:b}     
\includegraphics[width=0.8\columnwidth]{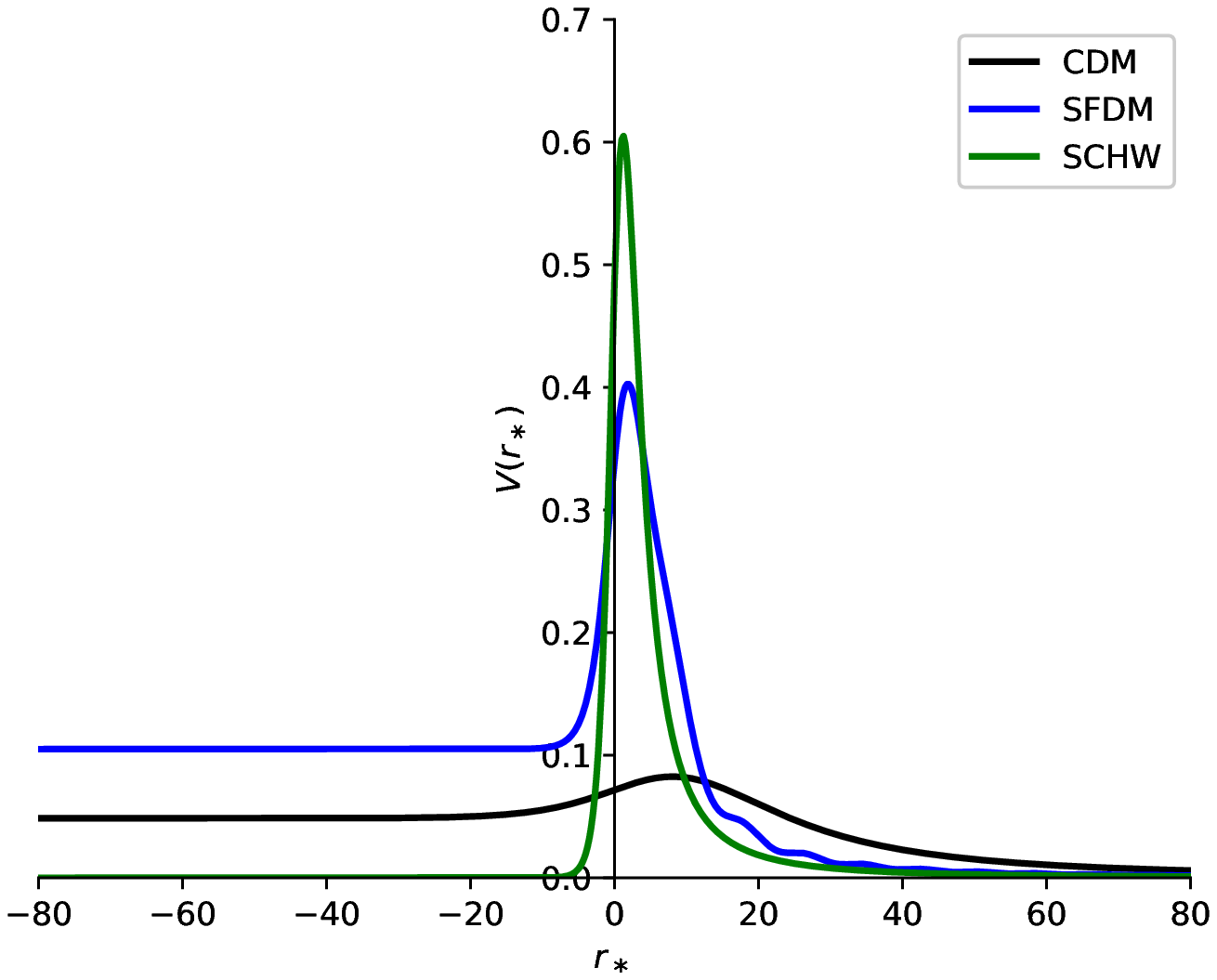}     
} 
{ 
\label{fig:b}     
\includegraphics[width=0.8\columnwidth]{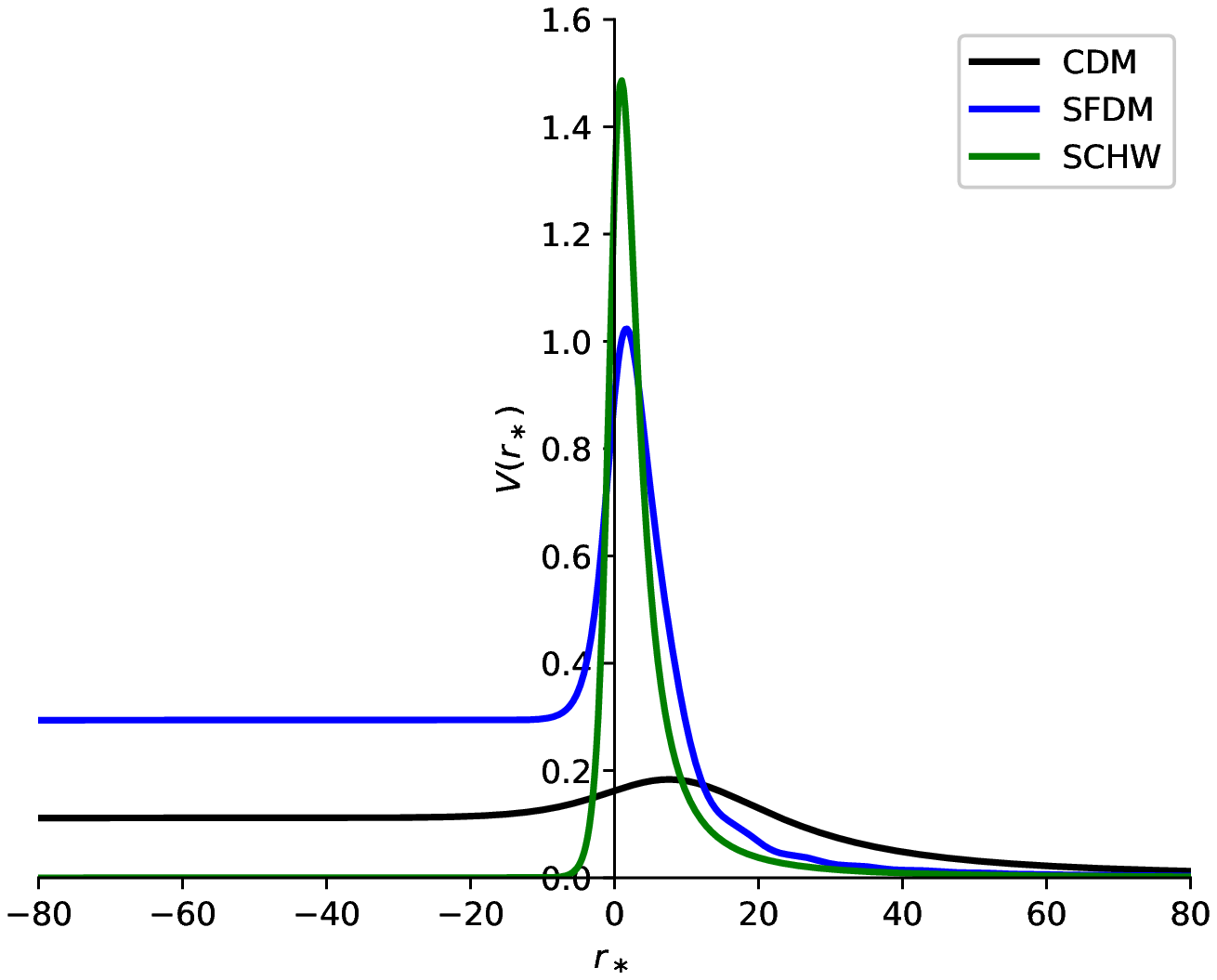}     
}
\caption{The effective potentials of gravitational field with the different space-times. The two panels, from left to right, are $l=2, 3$.}     
\label{fig:6}     
\end{figure*}

We introduce the tortoise coordinate; it has the  following form:
\begin{eqnarray}
dr_{\ast}=\frac{dr}{f(r)}.
\label{equ6}
\end{eqnarray}
\indent After separating the variables of Eqs.(\ref{equ4}) and (\ref{equ5}), a wavelike equation usually takes the following Schrödinger-like form for a stationary background:
\begin{equation}
-\frac{d^2\Psi}{d{r^2_*}}+V(r) \Psi=\omega ^{2}\Psi ,  
\label{equ7}                                                   
\end{equation}\\
and the effective potentials of CDM, SFDM and SCHW are as follows respectively:\\
\begin{widetext}
\begin{eqnarray}
V({\rm CDM})=((1+\frac{r}{R_{\rm c}})^{-\frac{8\pi \rho _{\rm c}R_{\rm c}^{3}}{r}}-\frac{2M}{r})[(\frac{2M }{r^{3}}+\frac{\Delta_{\rm CDM} }{r})(1-s^2)+\frac{l(l+1)}{r^{2}}],
\label{equ8}
\end{eqnarray}

\begin{eqnarray}
V({\rm SFDM})=[{\rm exp}(-\frac{8\rho _{\rm s}{R_{\rm s}^{2}}}{\pi }\frac{\sin(\pi r/R_{\rm s})}{\pi r/R_{\rm s}})-\frac{2M}{r}][(\frac{2M }{r^{3}}+\frac{\Delta_{\rm SFDM}}{r})(1-s^2)+\frac{l(l+1)}{r^{2}}],
\label{equ9}
\end{eqnarray}
\centerline{$\Delta _{\rm CDM} ={((1+\frac{r}{R_{\rm c}})^{-\frac{8\pi \rho _{\rm c}R_{\rm c}^{3}}{r}})}'=(1+\frac{r}{R_{\rm c}})^\frac{{-8\pi \rho_{\rm c}R_{\rm c}^{3}}}{r}(-\frac{8\pi R_{\rm c}^{2}\rho _{\rm c}}{r(1+\frac{r}{R_{\rm c}})}+\frac{8\pi R_{\rm c}^{3}\rho _{\rm c}\log(1+\frac{r}{R_{\rm c}})}{r^{2}})$,}\\

\centerline{$\Delta _{\rm SFDM}={{\rm exp}(-\frac{8\rho _{\rm s}{R_{\rm s}^{2}}}{\pi }\frac{\sin(\pi r/R_{\rm s})}{\pi r/R_{\rm s}})}'={\rm exp}(-\frac{8{R_{\rm s}^{3}}\rho _{\rm s}\sin(\frac{\pi r}{R_{\rm s}})}{\pi ^{2}r})(\frac{8{R_{\rm s}^{3}}\rho_{\rm s}\sin(\frac{\pi r}{R_{\rm s}})}{\pi ^{2}r^{2}}-\frac{8{R_{\rm s}^{2}}\rho_{\rm s}\cos(\frac{\pi r}{R_{\rm s}})}{\pi r})$.}
\begin{eqnarray}
V({\rm SCHW})=(1-\frac{2M}{r})(\frac{2M}{r^3}(1-s^2)+\frac{l(l+1)}{r^2}),
\label{equ10}
\end{eqnarray}
\end{widetext}
\indent Here, $\Delta$ is the first derivative of the transcendental term versus $r$, where $s=0$ corresponds to a scalar field and $s = 1$ corresponds to an electromagnetic field, and $l$ is the angular quantum number.\\
\indent The third panel in Figs.\ref{fig:1}-\ref{fig:3} show that the effective potentials of a Schwarzschild black hole increase with the increasing $l$, and decay at infinity, eventually disappear, then a black hole will be back in balance. Different from the Schwarzschild background, CDM and SFDM tend to a positive value at negative infinity but tend to $0$ at positive infinity. Figures \ref{fig:4}-\ref{fig:6} describe the effective potentials under the scalar field, the electromagnetic field and gravitational perturbation respectively. The maximum values of the effective potential in a dark matter halo are slightly less than that of the Schwarzschild black hole. Furthermore, one case(the first panel in Fig.\ref{fig:4}) shows that the effective potentials of SFDM have a positive and negative oscillation behavior when $r_*$ tends to positive infinity. With the increasing $l$, the oscillation behavior of effective potential weakens and its values tend to $0$.

\subsection{Gravitational perturbation of a BH in a dark matter halo(axial perturbations)} 
Gravitational perturbation, which means the metric perturbation, can be used to solve the perturbed problem of special black hole spacetime. Under this background, its perturbed components can be written as partial differential equations simply. These equations were originally given by Regge and Wheeler\cite{Regge}. As a sample, they calculated the simplest perturbed case for the Schwarzschild black hole.

Here,we will calculate the axial perturbations of the dark matter halo. First, we introduce the small perturbed term $h_{\mu \nu }$ to the background metric $\bar{g}_{\mu \nu }$. Then the resulting perturbed metric $g_{\mu \nu }$ can be written as
\begin{eqnarray}
g_{\mu \nu }=\bar{g}_{\mu \nu }+h_{\mu \nu }, \quad  and  \quad  h_{\mu \nu } \ll \bar{g}_{\mu \nu }.
\label{equ11}
\end{eqnarray}
With the perturbed metric, the perturbed Christoffel symbols also can be rewritten as
\begin{eqnarray}
\Gamma _{\mu \nu }^{\lambda  }=\bar{\Gamma }_{\mu \nu }^{\lambda  }+\delta \Gamma _{\mu \nu }^{\lambda  },
\label{equ12}
\end{eqnarray}
where $\bar{\Gamma }_{\mu \nu }^{\alpha }$ are Christoffel symbols and the $\delta \Gamma _{\mu \nu }^{\alpha }$ can be written as
\begin{eqnarray}
\delta \Gamma _{\mu \nu }^{\lambda  }=\frac{1}{2}\bar{g}^{\lambda  \beta }(h_{\mu \beta ;\nu +h_{\nu \beta ;\mu }-h_{\mu \nu ;\beta }}),
\label{equ13}
\end{eqnarray}
Then, the perturbed Ricci tensor can be written as 
\begin{eqnarray}
R_{\mu \nu }=\bar{R}_{\mu \nu }+\delta R_{\mu \nu },
\label{equ14}
\end{eqnarray}
where
\begin{eqnarray}
\delta R_{\mu \nu }=\delta \Gamma _{\mu \lambda  ;\nu  }^{\lambda}-\delta \Gamma _{\mu \nu;\lambda  }^{\lambda  },
\label{equ15}
\end{eqnarray}
and the symbol of $;\nu$ is the covariant derivative to the background metric $\bar{g}_{\mu \nu }$.

Due to the perturbed term $\delta R_{\mu \nu }$ of the Ricci tensor $R_{\mu \nu }$ has no contribution\cite{t.k}. So, the field equation of the axial perturbation can be written as
\begin{eqnarray}
\delta R_{\mu \nu }=0.
\label{equ16}
\end{eqnarray}

In the dark matter halo, we consider the case of CDM and SFDM. Because of the spherically symmetric metrics, we can introduce the odd perturbations to the $\bar{g}_{\mu \nu }$. The perturbed term $h_{\mu \nu}^{odd}$ can be written as \cite{Regge}
\begin{widetext}
\begin{eqnarray}
h_{\mu \nu }^{odd}=\begin{pmatrix}
0 &  0& 0 & h_{0}(t,r)\\ 
 0& 0 &  0& h_{1}(t,r)\\ 
0 & 0 & 0 &0 \\ 
h_{0}(t,r) &h_{1}(t,r)  &0  & 0
\end{pmatrix}{\rm sin}\theta \partial \theta P_{l}({\rm cos}\theta )
\label{equ17}
\end{eqnarray}
\end{widetext}
where $P_{l}({\rm cos}\theta )$ are the Legendre polynomials of order $l$.

The component forms of Eqs. (\ref{equ16}) can be written as follows:

\begin{widetext}
\begin{eqnarray}
\delta R_{t\varphi }=\frac{ \left(2 r f'(r)-l(l+1)\right) \text{h0}(t,r)}{2 r^2} +  \frac{f(r)}{2} \left(\frac{\partial ^2}{\partial r^2}h0(t,r)-\frac{2}{r}\frac{\partial }{\partial t}h1(t,r)-\frac{\partial ^2}{\partial r\partial t}h1(t,r)\right)=0,
\label{equ18}
\end{eqnarray}
\begin{eqnarray}
\delta R_{r\varphi }=\frac{ \left(2 r f'(r)-l(l+1)+2f(r)\right) \text{h1}(t,r)}{r^2} + \frac{1}{f(r)} \left(\frac{\partial^2 }{\partial r \partial t}h0(t,r)-\frac{\partial^2 }{\partial t^2}h1(t,r)-\frac{2}{r} \frac{\partial }{\partial t}h0(t,r)\right)=0,
\label{equ19}
\end{eqnarray}
\begin{eqnarray}
\delta R_{\theta \varphi }=f'(r) \text{h1}(t,r)-\frac{1}{f(r)}\frac{\partial }{\partial t}h0(t,r)+f(r) \frac{\partial }{\partial r}h1(t,r)=0,
\label{equ20}
\end{eqnarray}
\end{widetext}

where the $f'(r)$ denote $\frac{d}{dr}f(r)$. Then we can eliminate the term of $\frac{\partial }{\partial t}h0(t,r)$ in combination with Eqs. (\ref{equ19}) and (\ref{equ20})\cite{m.e,m.j,m.e,a.c,k.a,24,25,26}. We should define $\Psi (t,r)=\frac{f(r)}{r}h1(t,r)$. The resulting equation can be written as 
\begin{widetext}
\begin{eqnarray}
\frac{\partial^2 }{\partial t^2}\Psi -\frac{f}{r}\frac{\partial }{\partial r}[f\frac{\partial }{\partial r}(r\Psi )]+\frac{2f^2}{r^2}\frac{\partial }{\partial r}(r\Psi )+\frac{f[l(l+1)-2rf'-2f]}{r^2}\Psi =0.
\label{equ21}
\end{eqnarray}
\end{widetext}
Now, we use Eq.(\ref{equ6}) to perform coordinate transformation on Eq.(\ref{equ21}). The resulting equation can be rewritten as \\
\begin{widetext}
\begin{eqnarray}
\frac{\partial^2 }{\partial t^2}\Psi -\frac{\partial^2 }{\partial r_*^2}\Psi +V_{\rm G}(r)\Psi =0.
\label{equ22}
\end{eqnarray}
\end{widetext}
$V_{\rm G}(r)$ is the effective potential of gravitational perturbation. The following Eqs.(\ref{equ23} - \ref{equ25}) are respectively correspond to the case of CDM, SFDM and SCHW,
\begin{widetext}
\begin{eqnarray}
V_{{\rm G}}({\rm CDM})=\frac{\left(\left(\frac{r+\text{Rs}}{\text{Rs}}\right)^{-\frac{8 \pi  \rho  \text{Rs}^3}{r}}-\frac{2 M}{r}\right) \left(l (l+1) r-6 M+24 \pi  \rho  \text{Rs}^2
   \left(\frac{r+\text{Rs}}{\text{Rs}}\right)^{-\frac{8 \pi  \rho  \text{Rs}^3}{r}-1} \left(r-(r+\text{Rs}) \log \left(\frac{r+\text{Rs}}{\text{Rs}}\right)\right)\right)}{r^3}
\label{equ23}
\end{eqnarray}
\begin{eqnarray}
V_{{\rm G}}({\rm SFDM})=\frac{\left(e^{-\frac{8 \rho  R^3 \sin \left(\frac{\pi  r}{R}\right)}{\pi ^2 r}}-\frac{2 M}{r}\right) \left(l (l+1)-\frac{6 M}{r}+\frac{24 \rho  R^2 e^{-\frac{8 \rho  R^3 \sin \left(\frac{\pi 
   r}{R}\right)}{\pi ^2 r}} \left(\pi  r \cos \left(\frac{\pi  r}{R}\right)-R \sin \left(\frac{\pi  r}{R}\right)\right)}{\pi ^2 r}\right)}{r^2}
\label{equ24}
\end{eqnarray}
\begin{eqnarray}
V_{{\rm G}}({\rm SCHW})=\frac{(1-\frac{2M}{r})\left ( l(l+1)-6M/r \right )}{r^2}
\label{equ25}
\end{eqnarray}
\end{widetext}

\subsection{The WKB method}
When calculating QNM frequencies, we use the WKB method. This method was first proposed by Schutz and Will \cite{B}, and then promoted by Iyer, Will, and Konoplya \cite{S,S.,R}. In order to obtain the QNM frequencies, here, we use the sixth-order WKB formula, which has the following form:
\begin{eqnarray}
\frac{i(\omega ^{2}-V_{0})}{\sqrt{-2V_{0}^{''}}}-\sum ^{6}_{i=2}\Lambda _{i}=n+\frac{1}{2}, \quad(n=0,1,2, \cdots)    
\label{equ26}           
\end{eqnarray}
where $V_{0}$ is the maximum value of the effective potentials, $\Lambda _{i}$ is the $i$th order revision terms depending on the values of the effective potential, and more details can be found in Ref.\cite{R. A. Konoplya*}. In this expression, the WKB formula is related to the effective potential directly, then the effective potentials of the echo depends on these three parameters: $M$, $R$, and $\rho$. In addition, we get QNM frequencies when the number of overtone $n=0$. The WKB program is compiled with $Mathematica$ and can be obtained on the website\cite{http}.

\subsection{The finite difference method}
Equation (\ref{equ7}) is second order differential equation about space. We rewrite Eq. (\ref{equ7}) without implying the stationary ansatz ($\Phi \sim e^{-i\omega t}$) and it has the following form,
\begin{eqnarray}
-\frac{\partial ^{2}\Psi}{\partial t^{2}}+\frac{\partial ^2\Psi}{\partial {r^2_*}}-V(r)\Psi=0.     
\label{equ27}
\end{eqnarray}

Its form is similar to Eq.(\ref{equ22}) and there is no analytical solution to this equation presently. Here, we use the finite difference method first developed by Gundlach, Price, and Pullin \cite{Gundlach C} to analyze the dynamical evolutions of the equation. We introduce the light-cone coordinates $u=t-r_*$ and $v=t+r_*$; the wavelike equation can be written in the following form:
\begin{eqnarray}
-4\frac{\partial ^{2}\psi (\mu ,\nu )}{\partial \mu \partial \nu }=V(\mu ,\nu)\psi (\mu ,\nu ),
\label{equ28}
\end{eqnarray}
where $r_*$ is a tortoise coordinate, and it can be obtained by first-order approximation from Eq.(\ref{equ6}). Equation (\ref{equ22}) is directly related to the effective potential. Therefore, the integration grid recorded in Ref.\cite {Moderski R} can be discretized as\\
\begin{widetext}
\begin{eqnarray}  
\Psi(N)=\Psi(W)+\Psi(E)-\Psi(S)-h ^2\frac{V(W)\Psi(W)+V(E)\Psi(E)}{8}+O(h ^4),     
\label{equ29}                                             
\end{eqnarray}
\end{widetext}
\indent Here, $h$ is the grid cell scale. The letters of the integration grid are $ N=(u+h ,v+h)$, $W=(u+h,v)$, $E=(u,v +h)$ and $S = (u,v)$ respectively. The initial condition is the Gaussian wave packet \cite{Moderski R,Moderski R1,Moderski R2}, $\psi (\mu=\mu _{0} ,\nu )=A {\rm exp}[-(\nu -\nu _{0})^{2}/{\sigma ^{2}}]$, where, $A=1$, $\nu_0=10$ and $\sigma=3$. In this way, we can obtain the dynamical evolution of QNMs. Furthermore, we find that the QNMs are not dependent on Gaussian initial parameters. To extract QNM frequencies, we use the Prony method to fit a signal by superposition of damped exponents \cite{Berti E},\\
\begin{eqnarray}
\Psi(t)\simeq \sum ^p_{i=1}C_ie^{-i\omega_it}.                                   
\end{eqnarray}
\indent Although the contribution of all overtones is reflected in the values of QNMs, the contribution of higher overtones is usually neglectable \cite{Z Stuchilk}, because the signals of QNMs have been greatly approximated to the fundamental mode. So, the frequency can be extracted in this way from the values of QNMs.

\begin{table*}[t!]
\caption{The frequencies of a quasinormal mode in the scalar field.}
\begin{ruledtabular}
\begin{tabular}{cclllclllclllcclllclllclll}
& \multicolumn{12}{c}{WKB method}                                                      &  & \multicolumn{12}{c}{Prony method}                                                     \\ \hline
$l$ & \multicolumn{4}{c}{CDM} & \multicolumn{4}{c}{SFDM} & \multicolumn{4}{c}{SCHW} &  & \multicolumn{4}{c}{CDM} & \multicolumn{4}{c}{SFDM} & \multicolumn{4}{c}{SCHW} \\ \hline
0 & \multicolumn{4}{c}{0.083879 - 0.024010$i$}    & \multicolumn{4}{c}{$\cdots$}     & \multicolumn{4}{c}{0.220928 - 0.201638$i$}     &  & \multicolumn{4}{c}{0.068764 - 0.008065$i$}    & \multicolumn{4}{c}{0.232957 - 0.115914$i$}     & \multicolumn{4}{c}{0.221031 - 0.210330$i$}     \\
1 & \multicolumn{4}{c}{0.194343 - 0.025055$i$}    & \multicolumn{4}{c}{$\cdots$}     & \multicolumn{4}{c}{0.586124 - 0.195422$i$}     &  & \multicolumn{4}{c}{0.194080 - 0.024921$i$}    & \multicolumn{4}{c}{0.331336 - 0.077789$i$}     & \multicolumn{4}{c}{0.586728 - 0.194592$i$}     \\
2 & \multicolumn{4}{c}{0.324268 - 0.025687$i$}    & \multicolumn{4}{c}{$\cdots$}     & \multicolumn{4}{c}{0.967955 - 0.201120$i$}     &  & \multicolumn{4}{c}{0.324206 - 0.025376$i$}    & \multicolumn{4}{c}{0.823326 - 0.105189$i$}     & \multicolumn{4}{c}{0.970030 - 0.191738$i$} \\
\end{tabular}
\end{ruledtabular}
\label{tab:1}
\end{table*}

\begin{table*}[t!]
\caption{The frequencies of a quasinormal mode in the electromagnetic field.}
\begin{ruledtabular}
\begin{tabular}{cclllclllclllcclllclllclll}
 & \multicolumn{12}{c}{WKB method}                                                      &  & \multicolumn{12}{c}{Prony method}                                                     \\ \hline
$l$ & \multicolumn{4}{c}{CDM} & \multicolumn{4}{c}{SFDM} & \multicolumn{4}{c}{SCHW} &  & \multicolumn{4}{c}{CDM} & \multicolumn{4}{c}{SFDM} & \multicolumn{4}{c}{SCHW} \\ \hline
1 & \multicolumn{4}{c}{0.181004 - 0.0250247$i$}    & \multicolumn{4}{c}{$\cdots$ }     & \multicolumn{4}{c}{0.496467 - 0.184438$i$}     &  & \multicolumn{4}{c}{0.177395 - 0.0248828$i$}    & \multicolumn{4}{c}{0.432891 - 0.103297$i$}     & \multicolumn{4}{c}{0.497133 - 0.184453$i$}     \\
2 & \multicolumn{4}{c}{0.316377 - 0.0256982$i$}    & \multicolumn{4}{c}{$\cdots$ }     & \multicolumn{4}{c}{0.915951 - 0.198093$i$}     &  & \multicolumn{4}{c}{0.314485 - 0.0253821$i$}    & \multicolumn{4}{c}{0.776575 - 0.105092$i$}     & \multicolumn{4}{c}{0.917546 - 0.188421$i$} \\
\end{tabular}
\end{ruledtabular}
\label{tab:2}
\end{table*}

\begin{table*}[t!]
\caption{The frequencies of a quasinormal mode in the gravitational perturbation.}
\begin{ruledtabular}
\begin{tabular}{cclllclllclllcclllclllclll}
 & \multicolumn{12}{c}{WKB method}                                                      &  & \multicolumn{12}{c}{Prony method}                                                     \\ \hline
$l$ & \multicolumn{4}{c}{CDM} & \multicolumn{4}{c}{SFDM} & \multicolumn{4}{c}{SCHW} &  & \multicolumn{4}{c}{CDM} & \multicolumn{4}{c}{SFDM} & \multicolumn{4}{c}{SCHW} \\ \hline
2 & \multicolumn{4}{c}{0.284007 - 0.0258836$i$}    & \multicolumn{4}{c}{$\cdots$ }     & \multicolumn{4}{c}{0.747107 - 0.178248$i$}     &  & \multicolumn{4}{c}{0.284456 - 0.0244245$i$}    & \multicolumn{4}{c}{0.425210 - 0.099847$i$}     & \multicolumn{4}{c}{0.748738 - 0.176901$i$}     \\
3 & \multicolumn{4}{c}{0.425956 - 0.0254432$i$}    & \multicolumn{4}{c}{$\cdots$ }     & \multicolumn{4}{c}{1.199220 - 0.189856$i$}     &  & \multicolumn{4}{c}{0.426723 - 0.0252413$i$}    & \multicolumn{4}{c}{0.827319 - 0.160441$i$}     & \multicolumn{4}{c}{1.203780 - 0.182989$i$} \\
\end{tabular}
\end{ruledtabular}
\label{tab:3}
\end{table*}

\section{Quasinormal modes of a black hole in a dark matter halo} \label{sec:3}
The dynamical evolutions of the QNMs are the solution of Eq.(\ref{equ28}). To make our calculations simply, we employ toy models to calculate the dynamical evolutions of QNMs for CDM and SFDM. First, we set $ M=0.5 $, and for CDM: $ \rho_{\rm c} = 0.001, R_{\rm c} = 6 $; For SFDM: $ \rho_{\rm s} = 0.01, R_{\rm s} = 3 $. Then, we study the cases of the scalar field, electromagnetic field and gravitational perturbation respectively, the situations with different $l$, and make comparisons with the Schwarzschild black hole. The QNMs are directly related to the effective potential. Figures \ref{fig:1}-\ref{fig:3} show that the effective potentials increase with the increasing $l$. We know that the QNM frequencies in a dark matter halo are related to dynamical evolutions and effective potentials. Generally speaking, the WKB method and the Prony method can be used to calculate the QNM frequencies. However, due to the effective potentials of SFDM having more than one peak value, the WKB method cannot be applied \cite{Z Stuchilk}. So the data of the WKB method in Tables \ref{tab:1}-\ref{tab:3} are no corresponding calculating results.

\indent Here, we use the sixth-order WKB method and the Prony method to calculate the values of frequencies under scalar field, electromagnetic field and gravitational perturbation, thereby obtaining the values in the Tables.

\begin{figure*}
\centering 
\subfigure[CDM]{
\label{fig:a}     
\includegraphics[width=0.6\columnwidth]{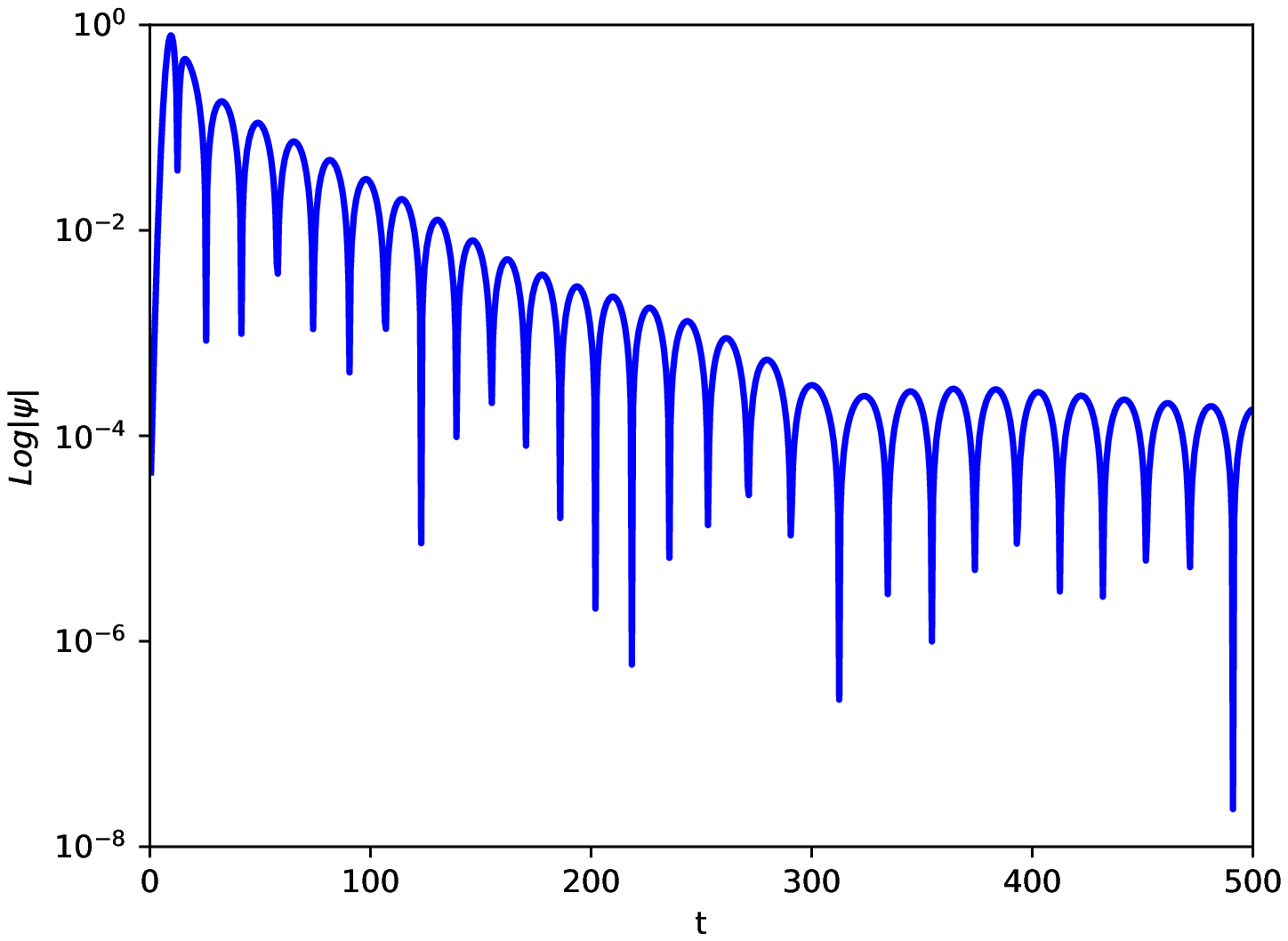}  
}     
\subfigure[SFDM]{  
                                                
\label{fig:b}     
\includegraphics[width=0.6\columnwidth]{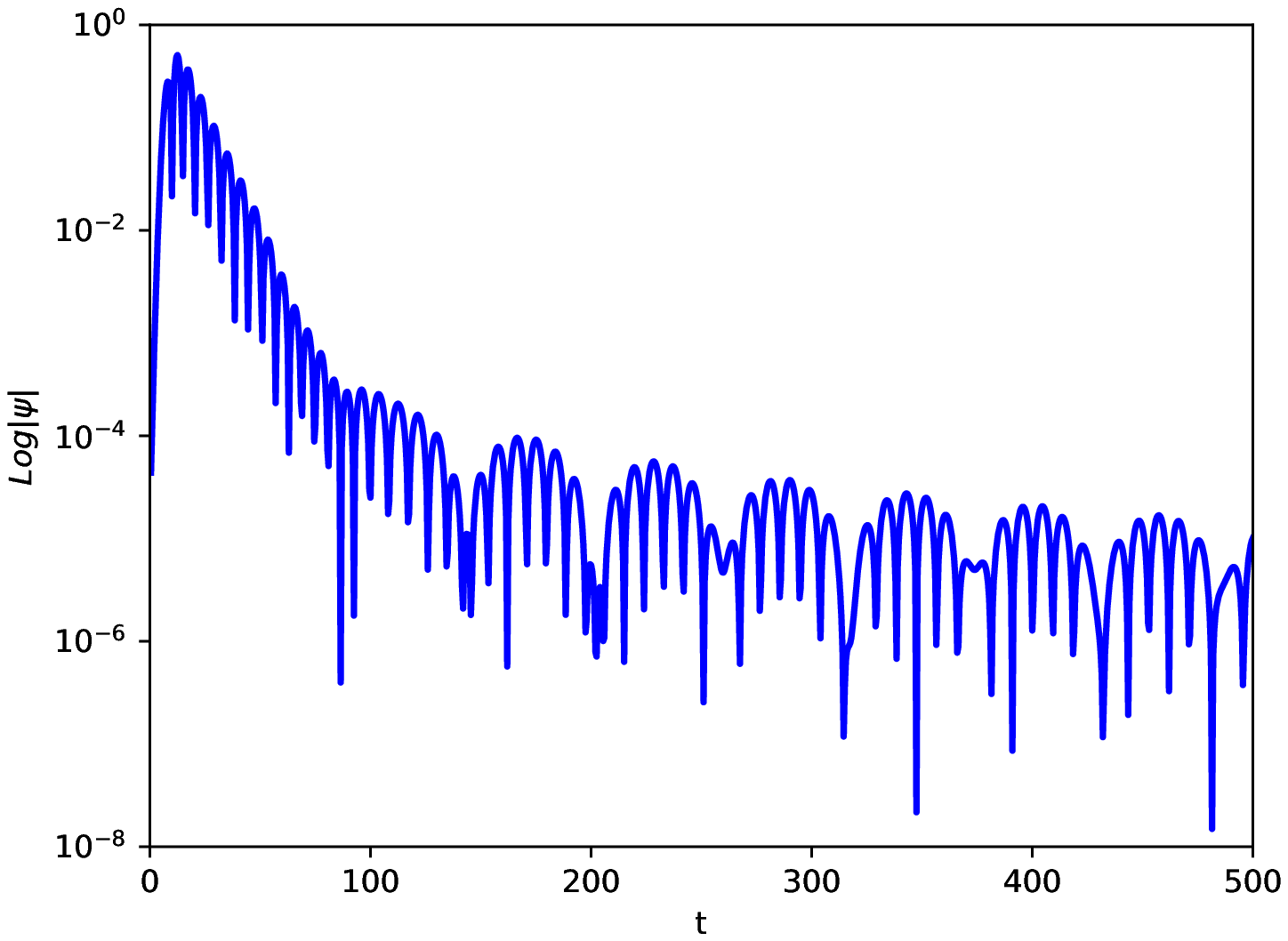}
}  
\subfigure[SCHW] {                                                 
\label{fig:b}     
\includegraphics[width=0.6\columnwidth]{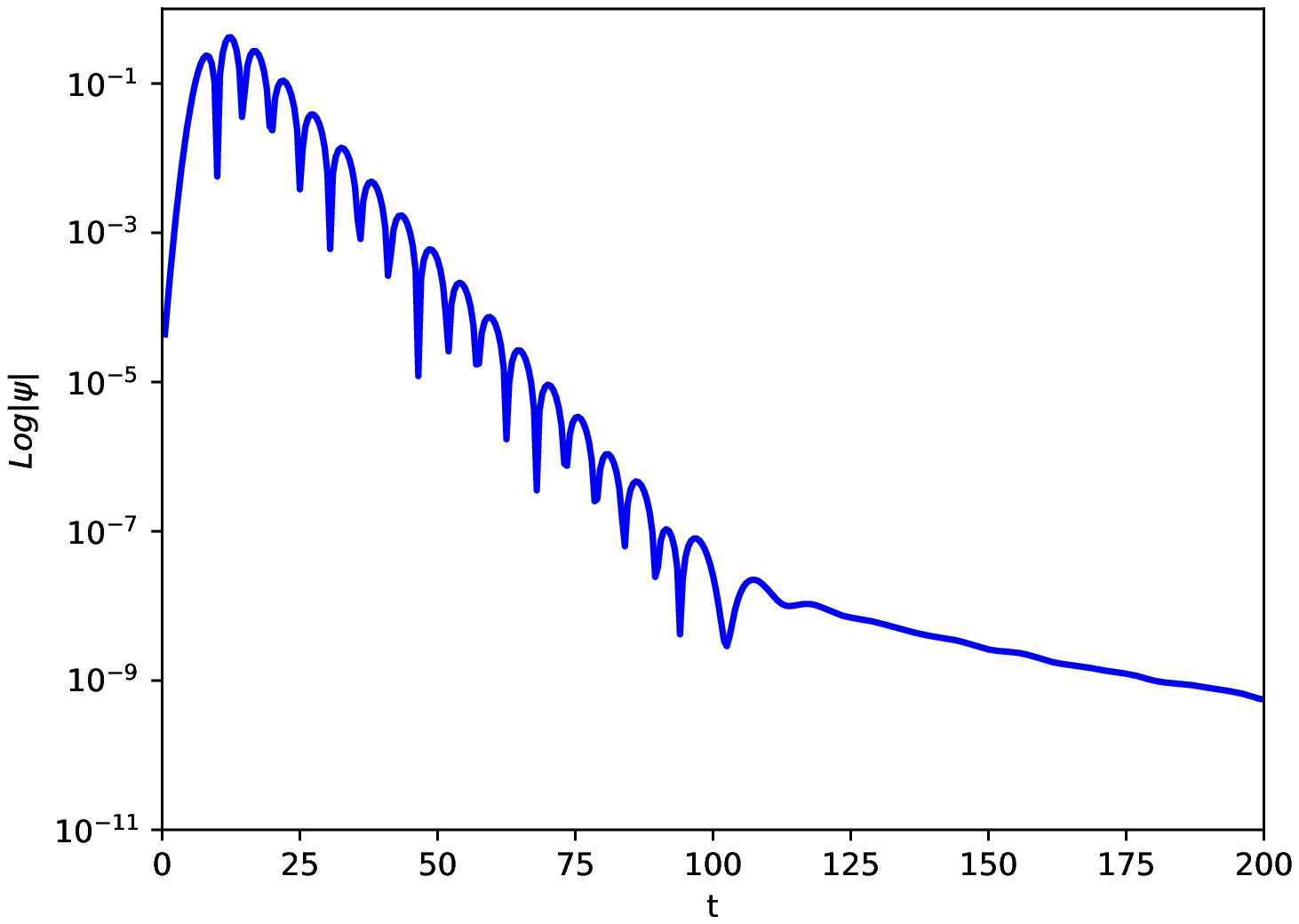} 

}   
\caption{The dynamical evolutions in the scalar perturbation with the different space-time ($M=0.5$, $l=1$, $s=0$).}     
\label{fig:7}     
\end{figure*}

\begin{figure*}
\centering 
\subfigure[CDM]{
\label{fig:a}     
\includegraphics[width=0.6\columnwidth]{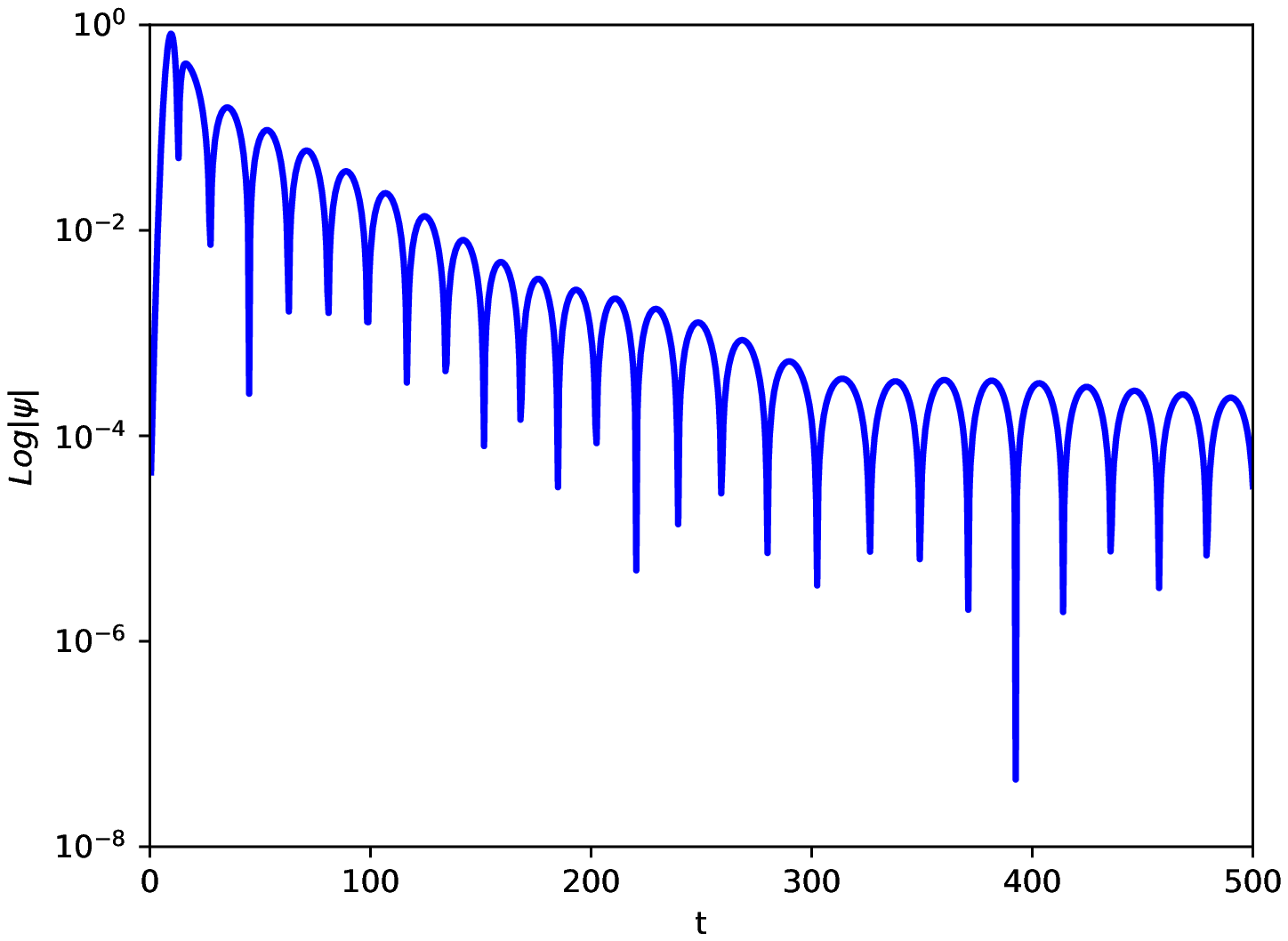}  
}     
\subfigure[SFDM]{  
                                                
\label{fig:b}     
\includegraphics[width=0.6\columnwidth]{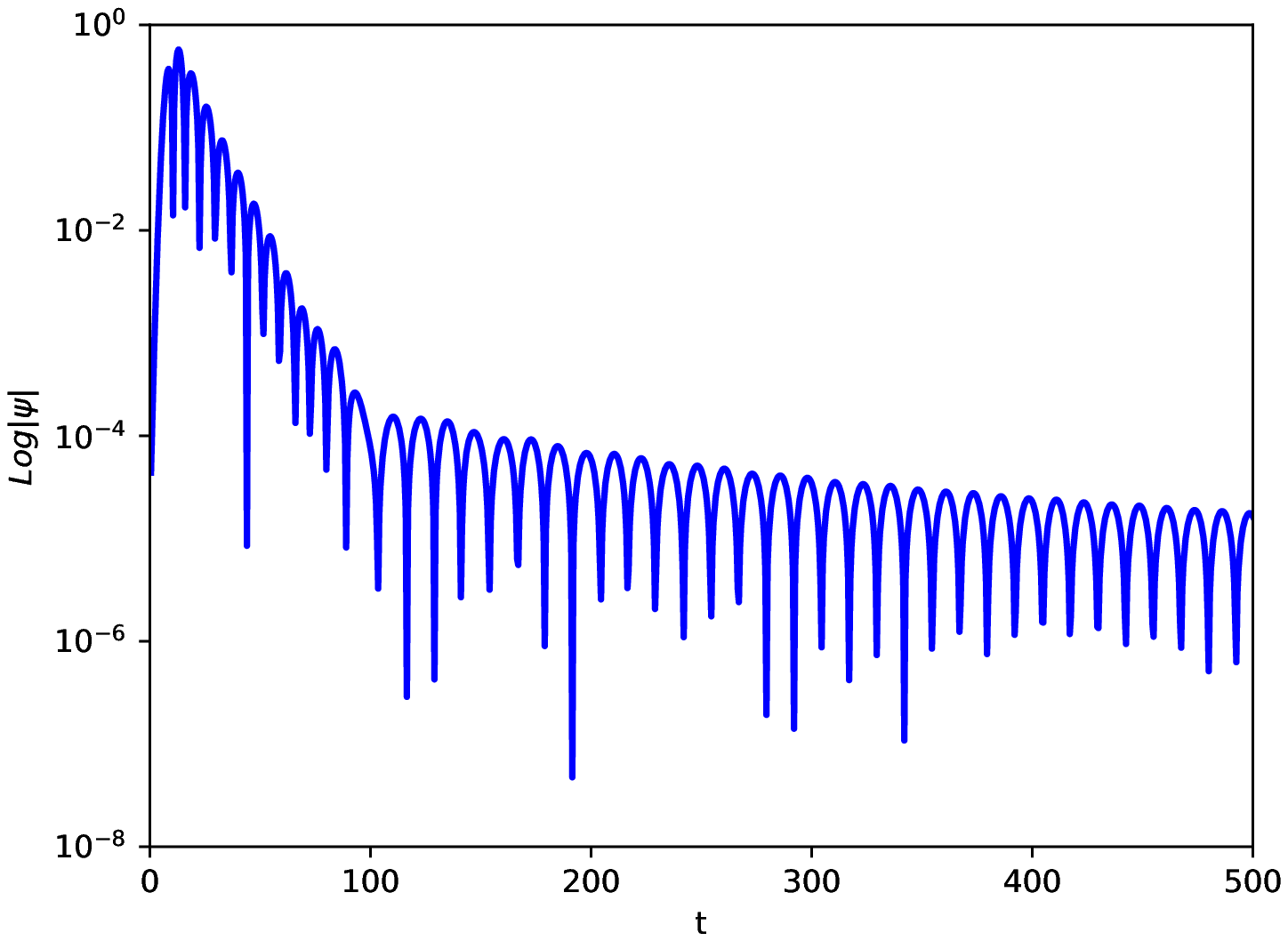}
}  
\subfigure[SCHW] {                                                 
\label{fig:b}     
\includegraphics[width=0.6\columnwidth]{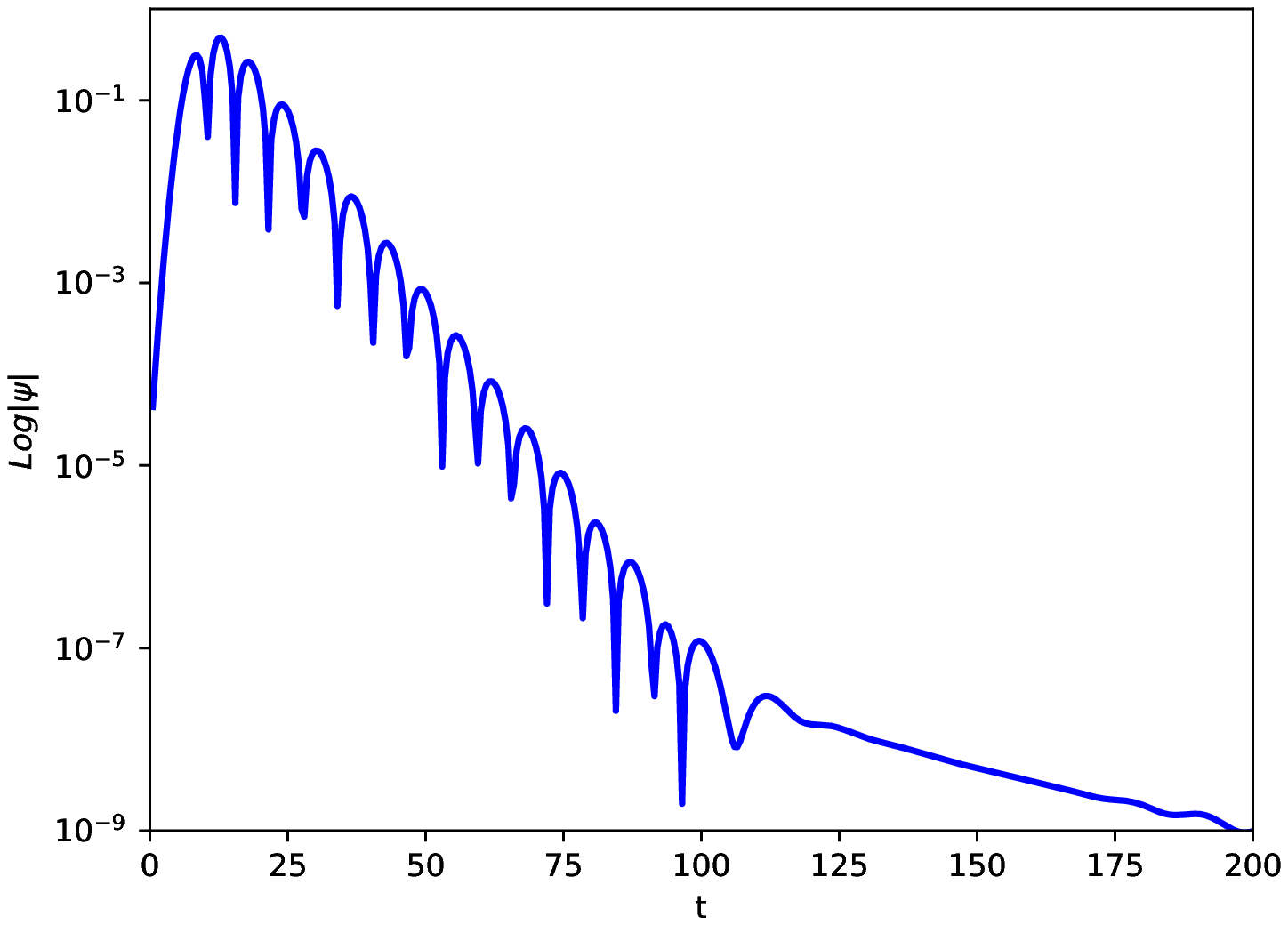} 
}   
\caption{The dynamical evolutions in the electromagnetic field with the different space-time ($M=0.5$, $l=1$, $s=1$).}     
\label{fig:8}     
\end{figure*}

\begin{figure*}
\centering 
\subfigure[CDM]{
\label{fig:a}     
\includegraphics[width=0.6\columnwidth]{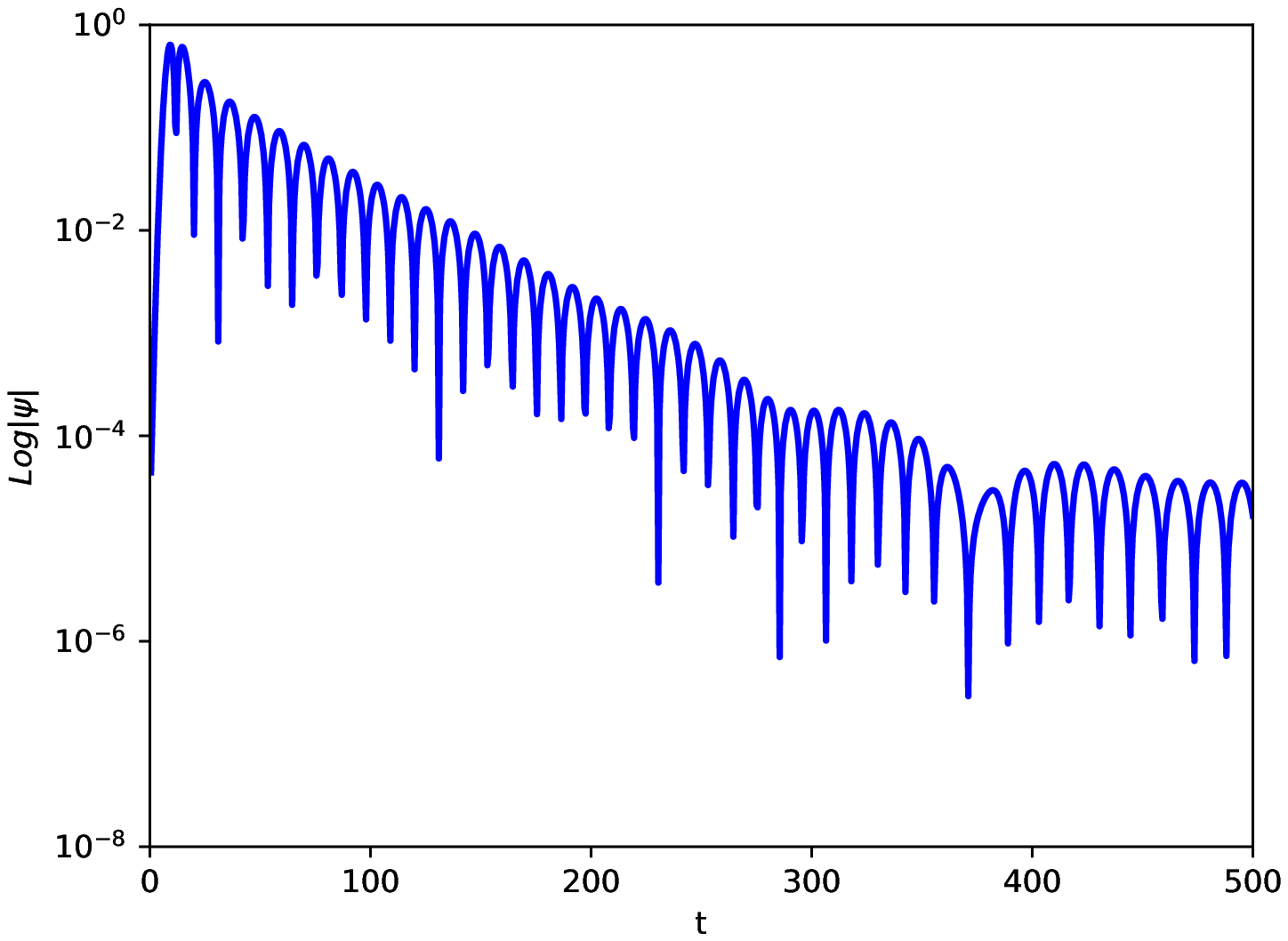}  
}     
\subfigure[SFDM]{  
                                                
\label{fig:b}     
\includegraphics[width=0.6\columnwidth]{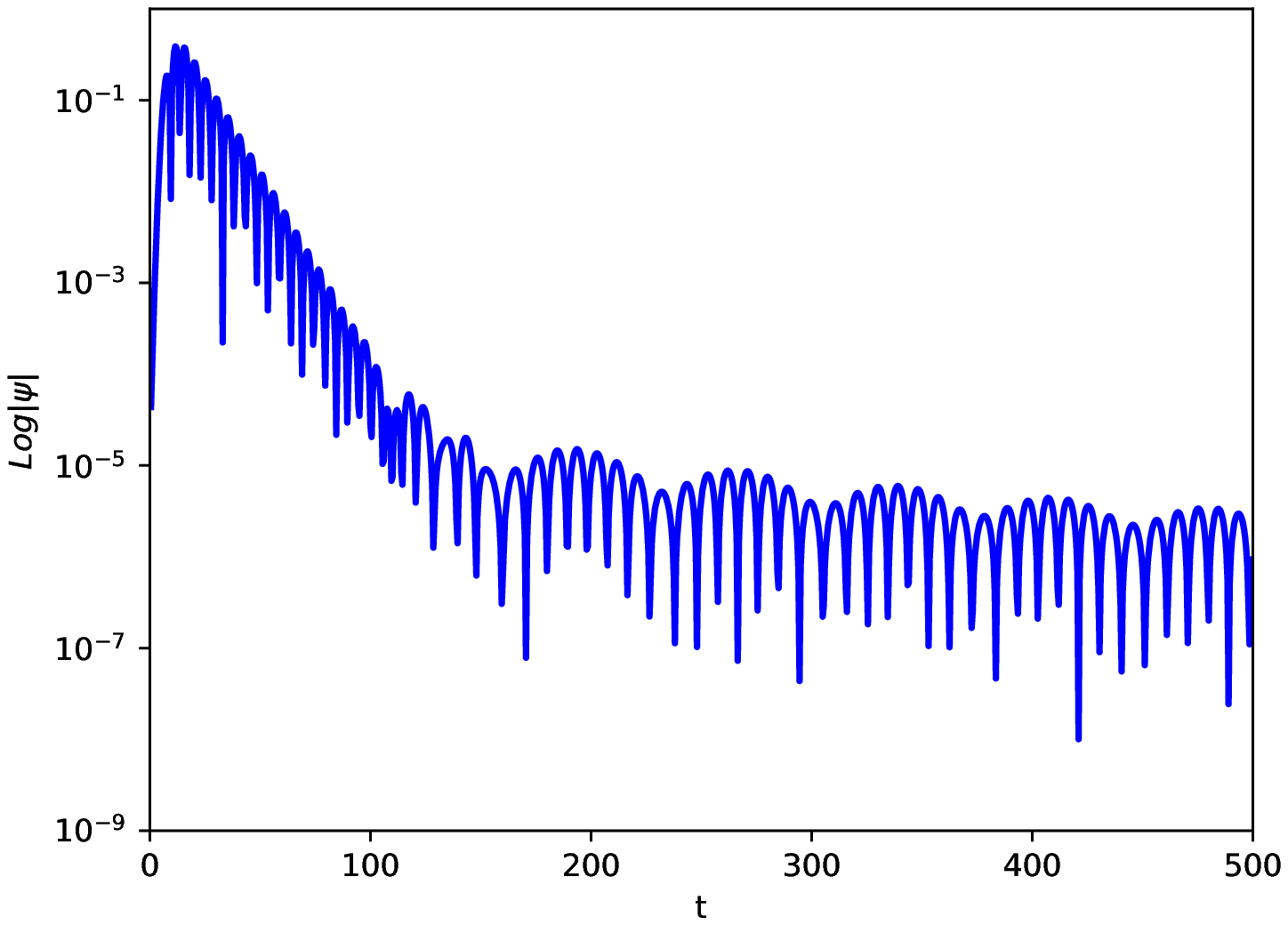}
}  
\subfigure[SCHW] {                                                 
\label{fig:b}     
\includegraphics[width=0.6\columnwidth]{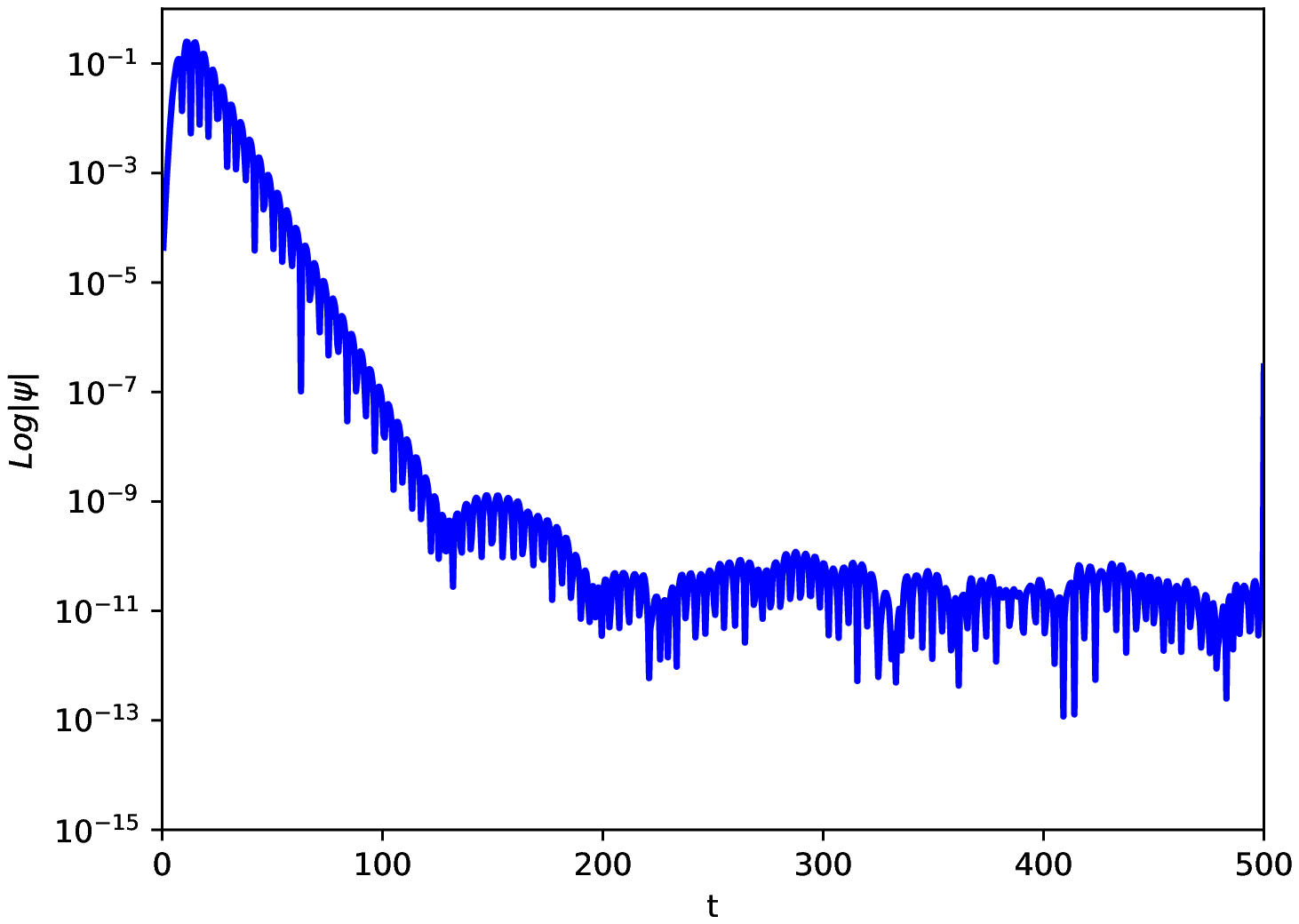} 

}   
\caption{The dynamical evolutions in the gravitational perturbation with the different space-time ($M=0.5$, $l=2$).}     
\label{fig:9}     
\end{figure*}

\begin{figure*}
\centering
\subfigure[CDM]{
\label{fig:a}     
\includegraphics[width=0.6\columnwidth]{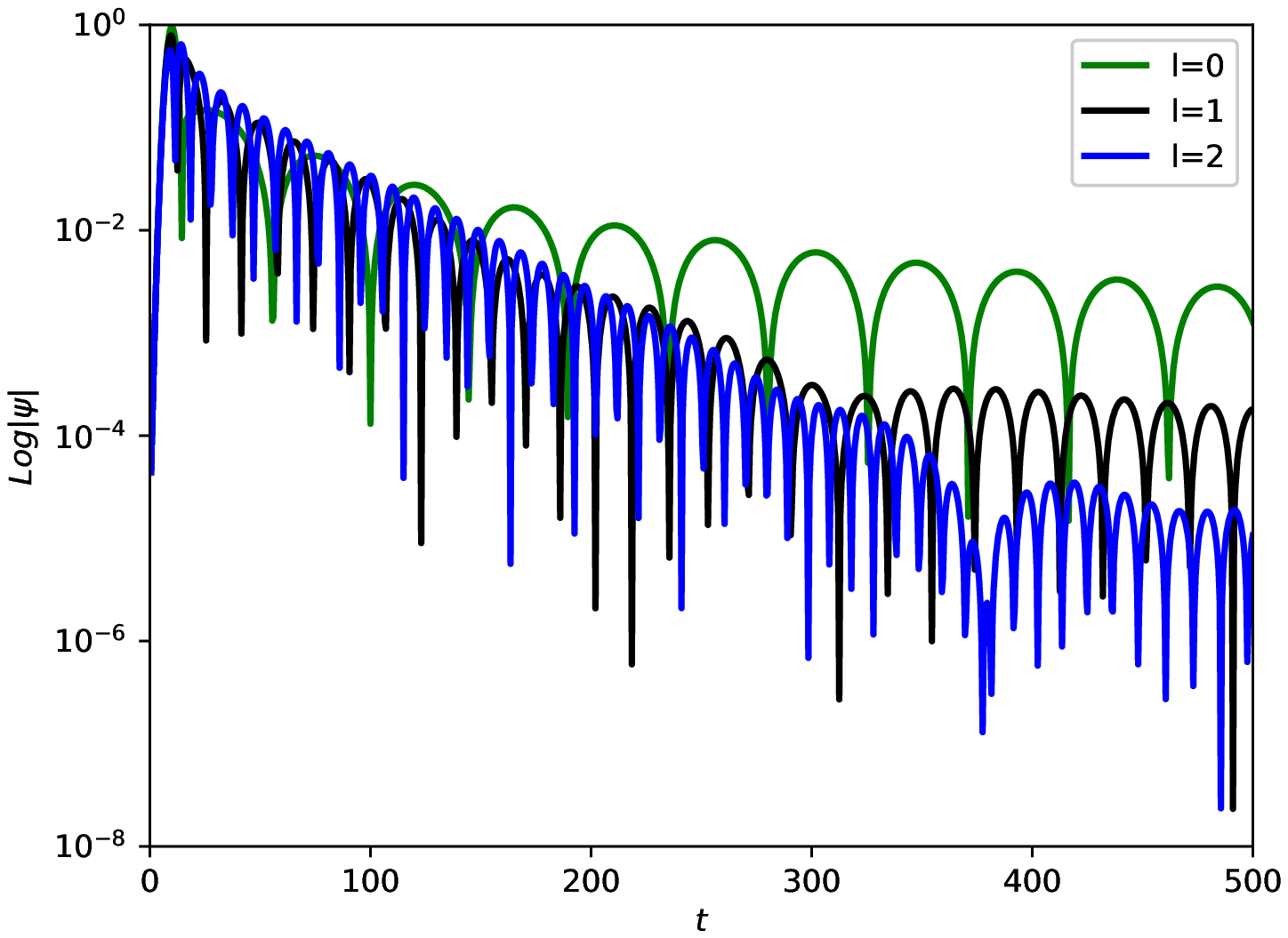}  
}     
\subfigure[SFDM]{                                                  
\label{fig:b}     
\includegraphics[width=0.6\columnwidth]{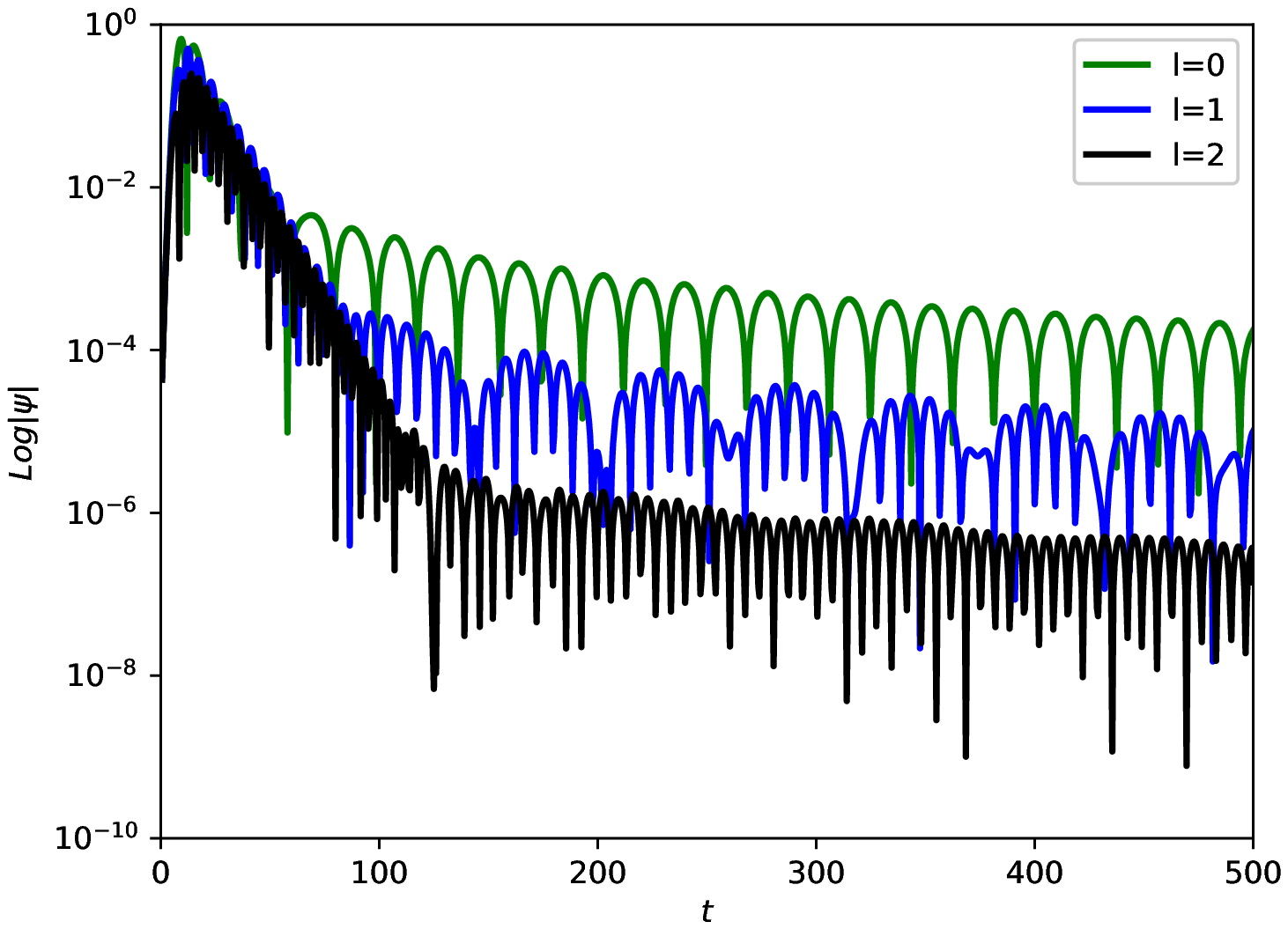}     
}    
\subfigure[SCHW]{ 
\label{fig:b}     
\includegraphics[width=0.6\columnwidth]{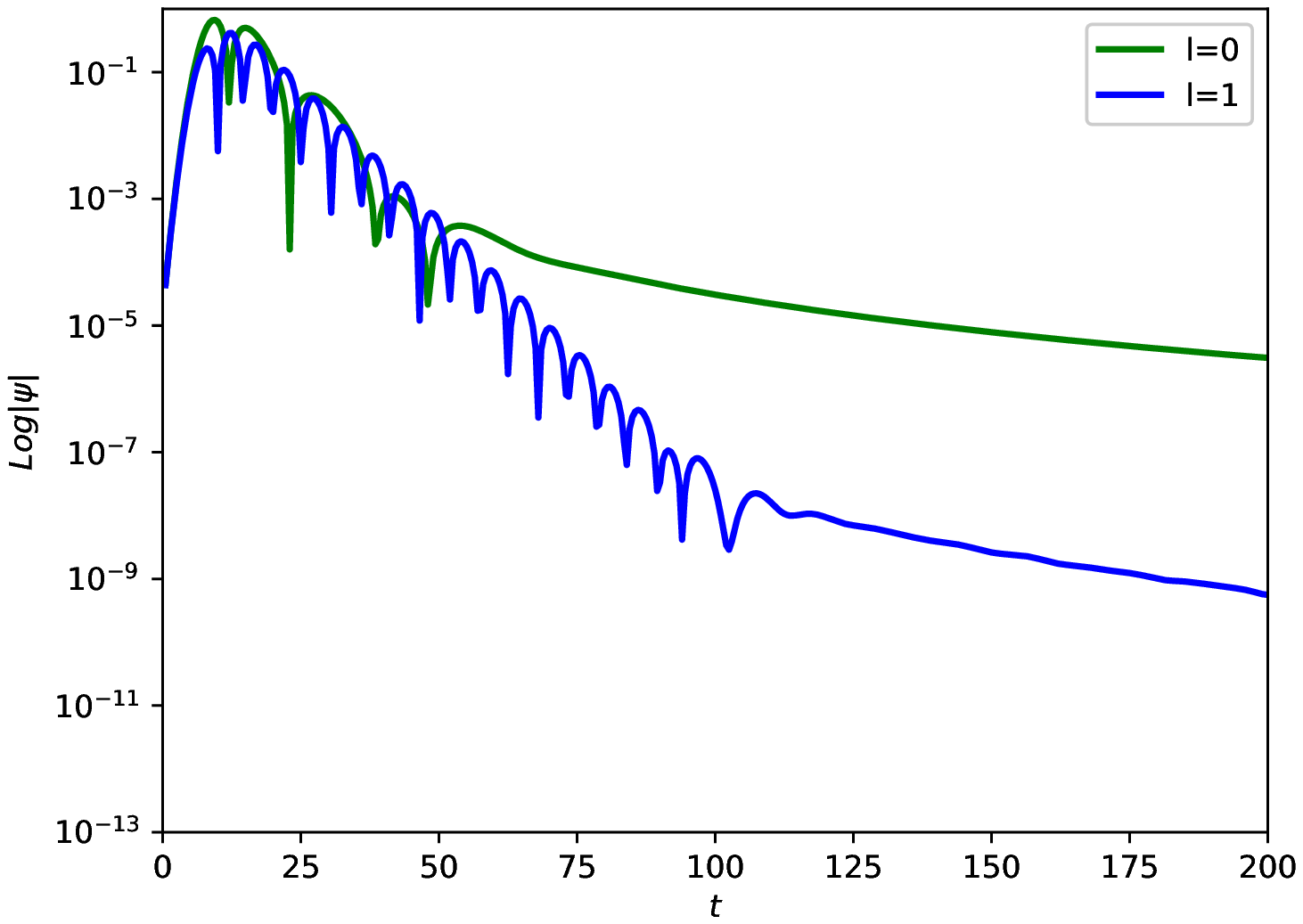}     
}   
\caption{Quasinormal modes in the scalar field with the different $l$. The ringing time of the QNMs increases with increasing parameter $l$ in each panel.  The parameters we used: $M=0.5$, $\nu _{0}=10$, $\sigma =3.$ }     
\label{fig:10}     
\end{figure*}

\begin{figure*}
\centering 
\subfigure[CDM]{
\label{fig:a}     
\includegraphics[width=0.6\columnwidth]{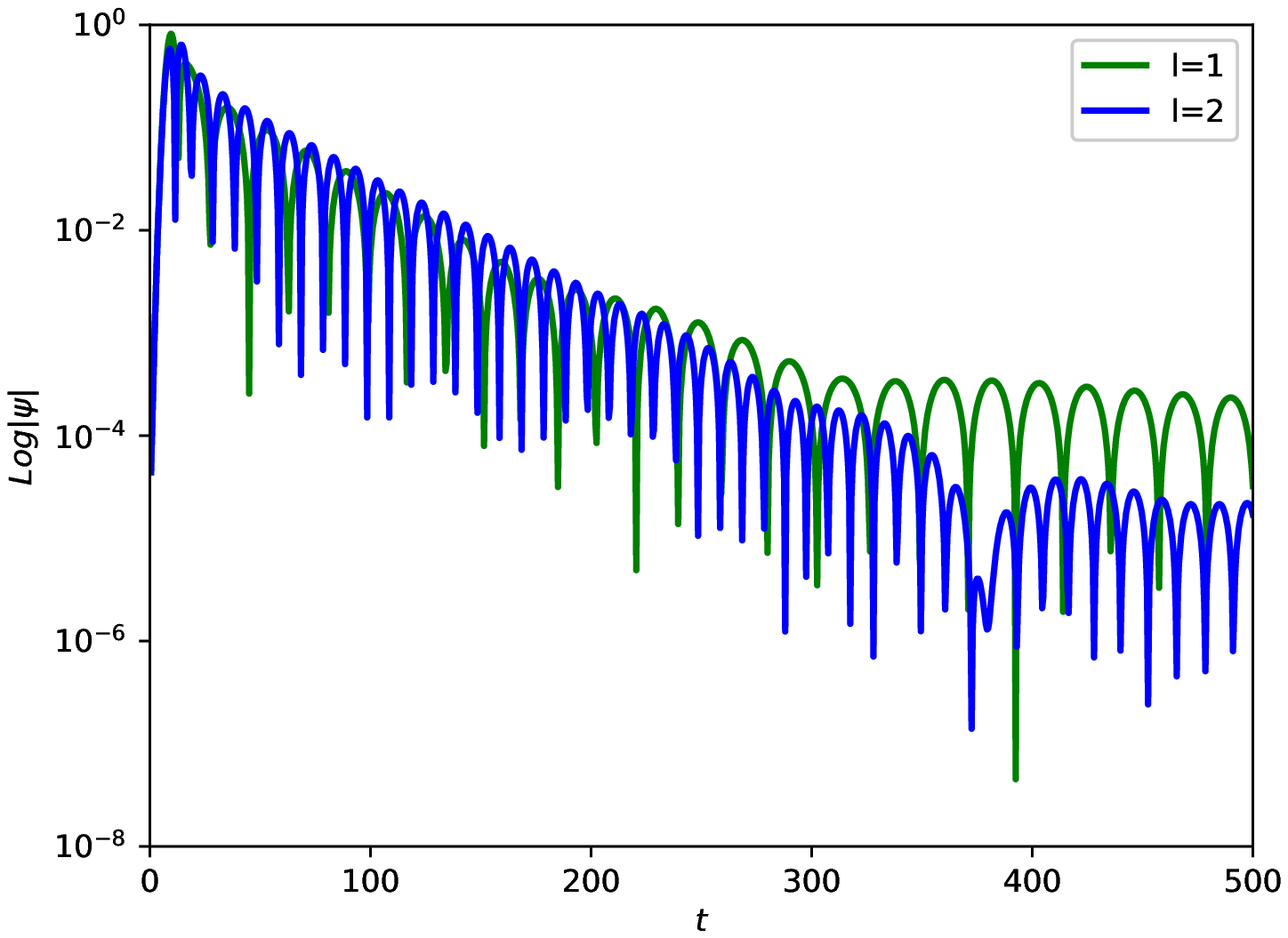}  
}     
\subfigure[SFDM]{                                                  
\label{fig:b}     
\includegraphics[width=0.6\columnwidth]{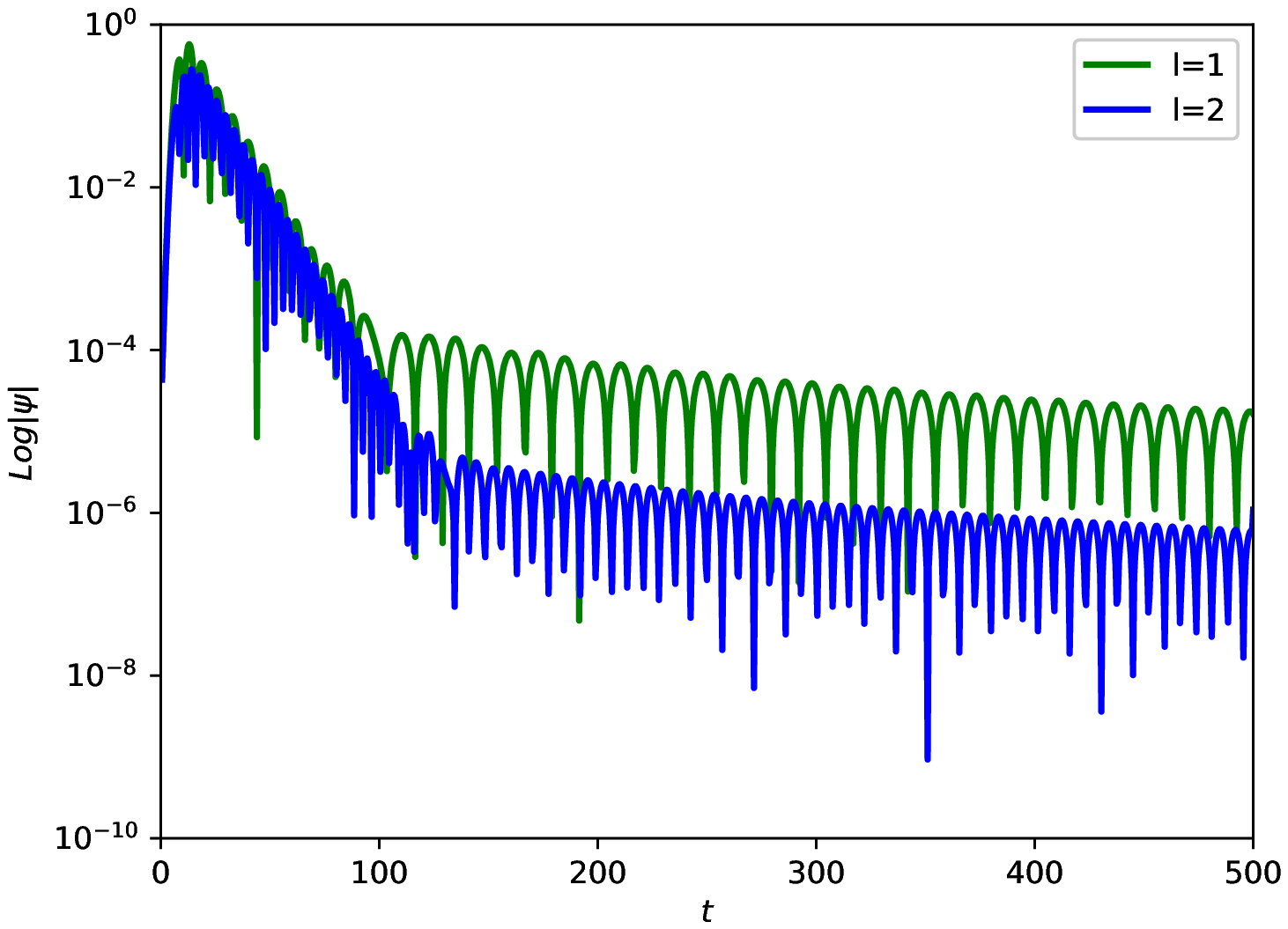}     
}    
\subfigure[SCHW]{ 
\label{fig:b}     
\includegraphics[width=0.6\columnwidth]{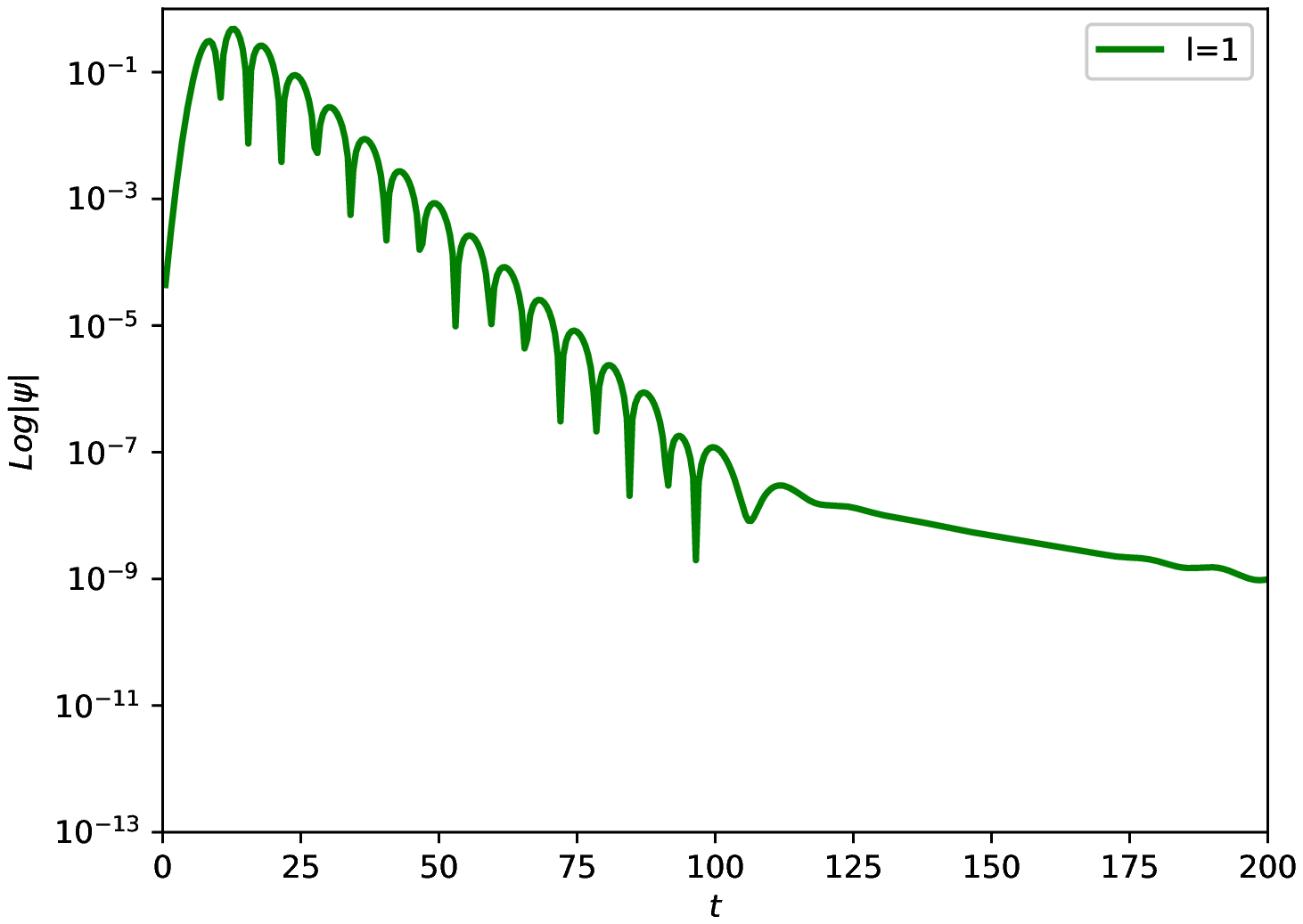}     
}   
\caption{Quasinormal modes in the electromagnetic field with the different $l$. The time of QNMs ringing increases with increasing parameter $l$ in each panel.}     
\label{fig:11}     
\end{figure*}

\begin{figure*}
\centering   
{                                                 
\label{fig:b}     
\includegraphics[width=0.6\columnwidth]{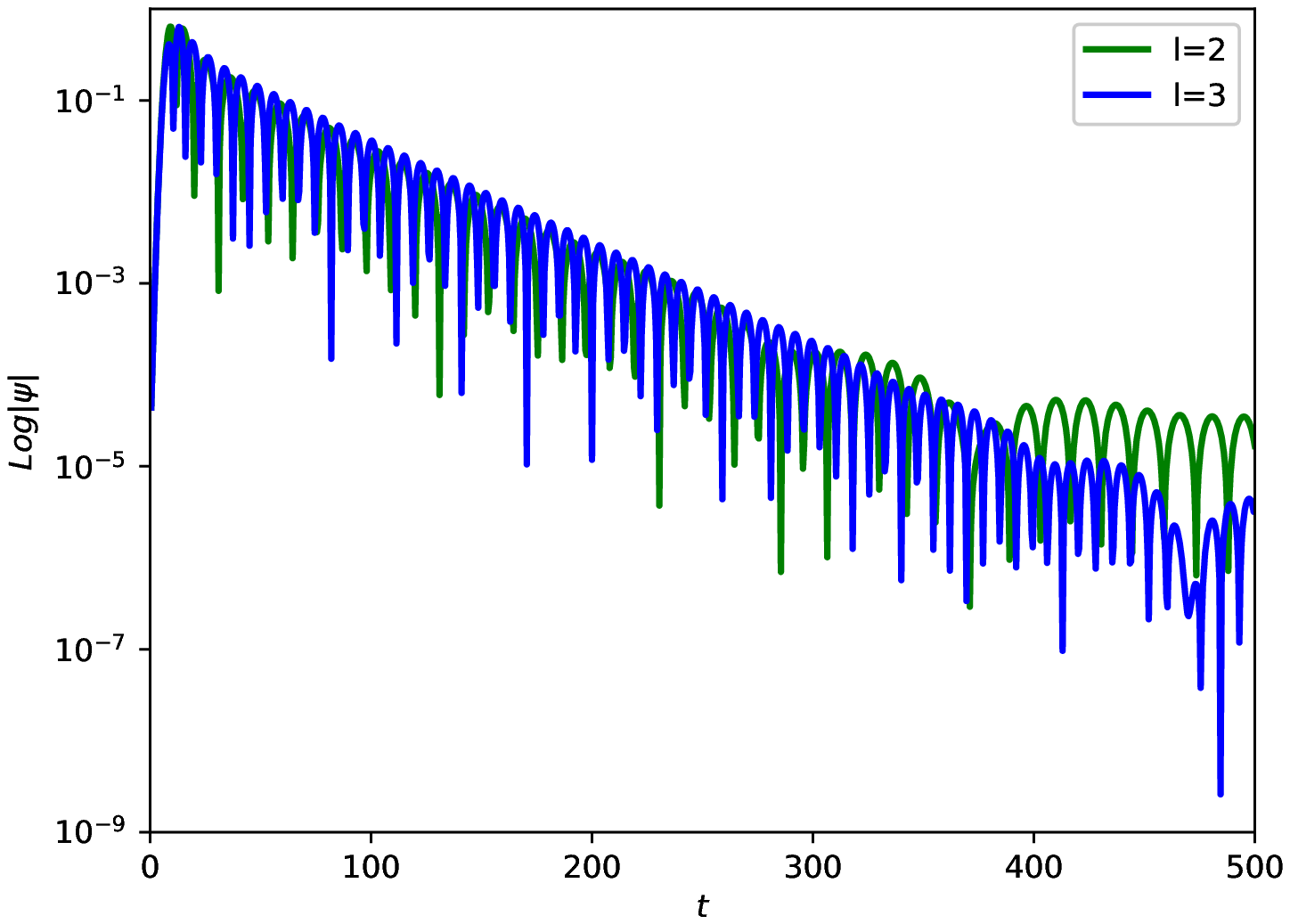}     
} 
{                                                
\label{fig:b}     
\includegraphics[width=0.6\columnwidth]{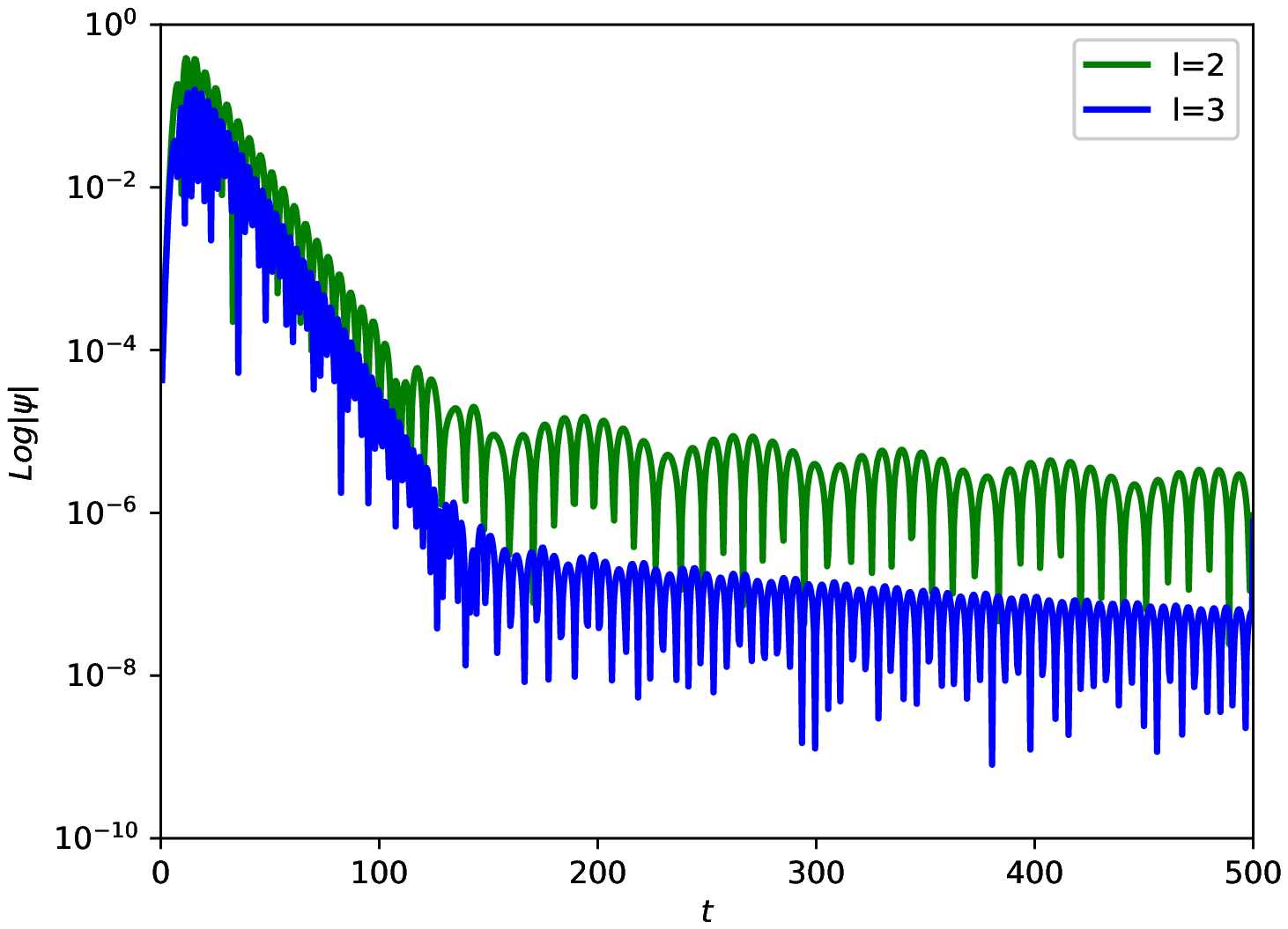}     
}    
{ 
\label{fig:b}     
\includegraphics[width=0.6\columnwidth]{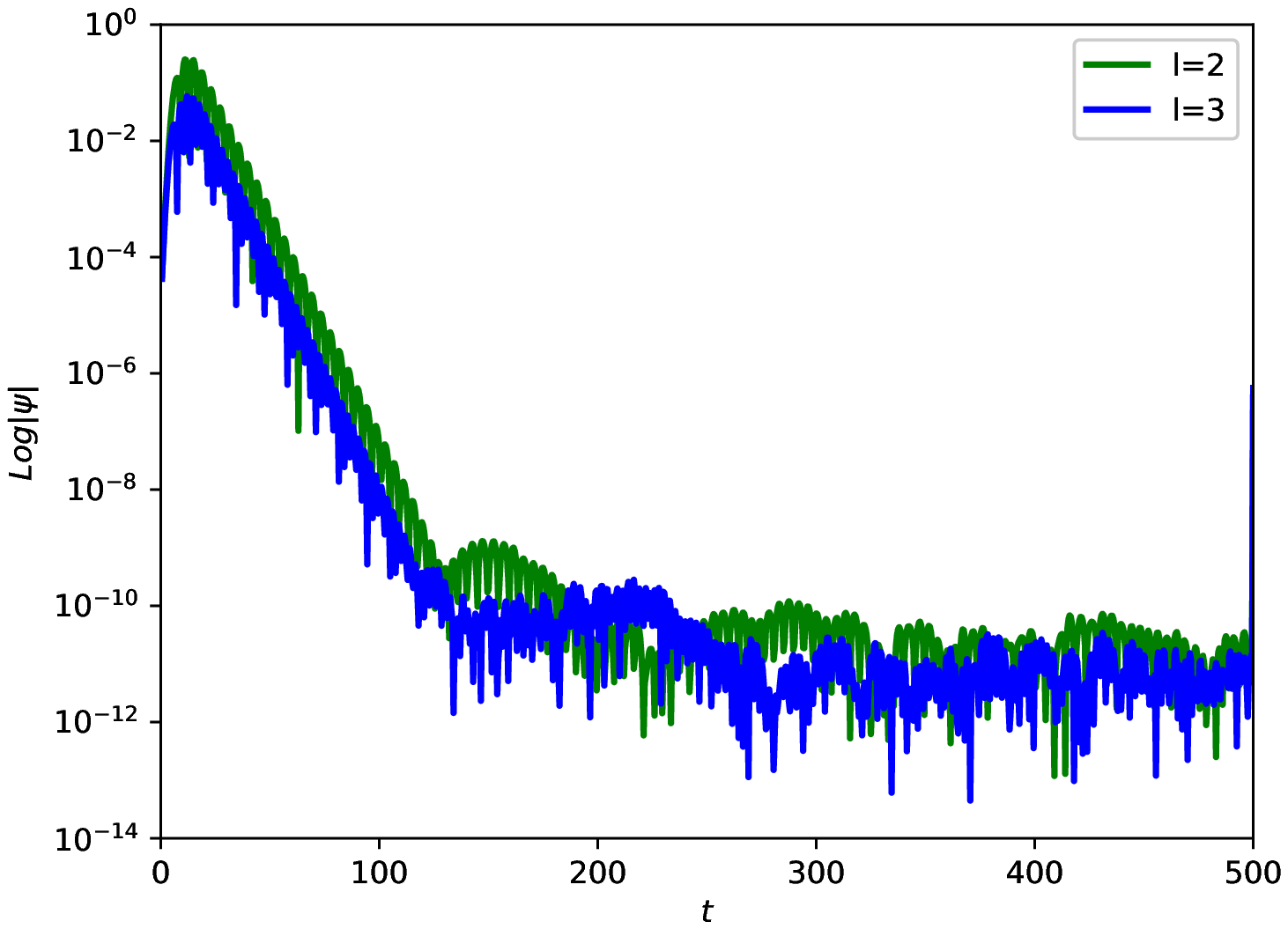}     
}   
\caption{Quasinormal modes in the gravitational perturbation with the different $l$.}     
\label{fig:12}   
\end{figure*}

\begin{figure*}
\centering   
\subfigure[$l=0$]{                                                 
\label{fig:b}     
\includegraphics[width=0.6\columnwidth]{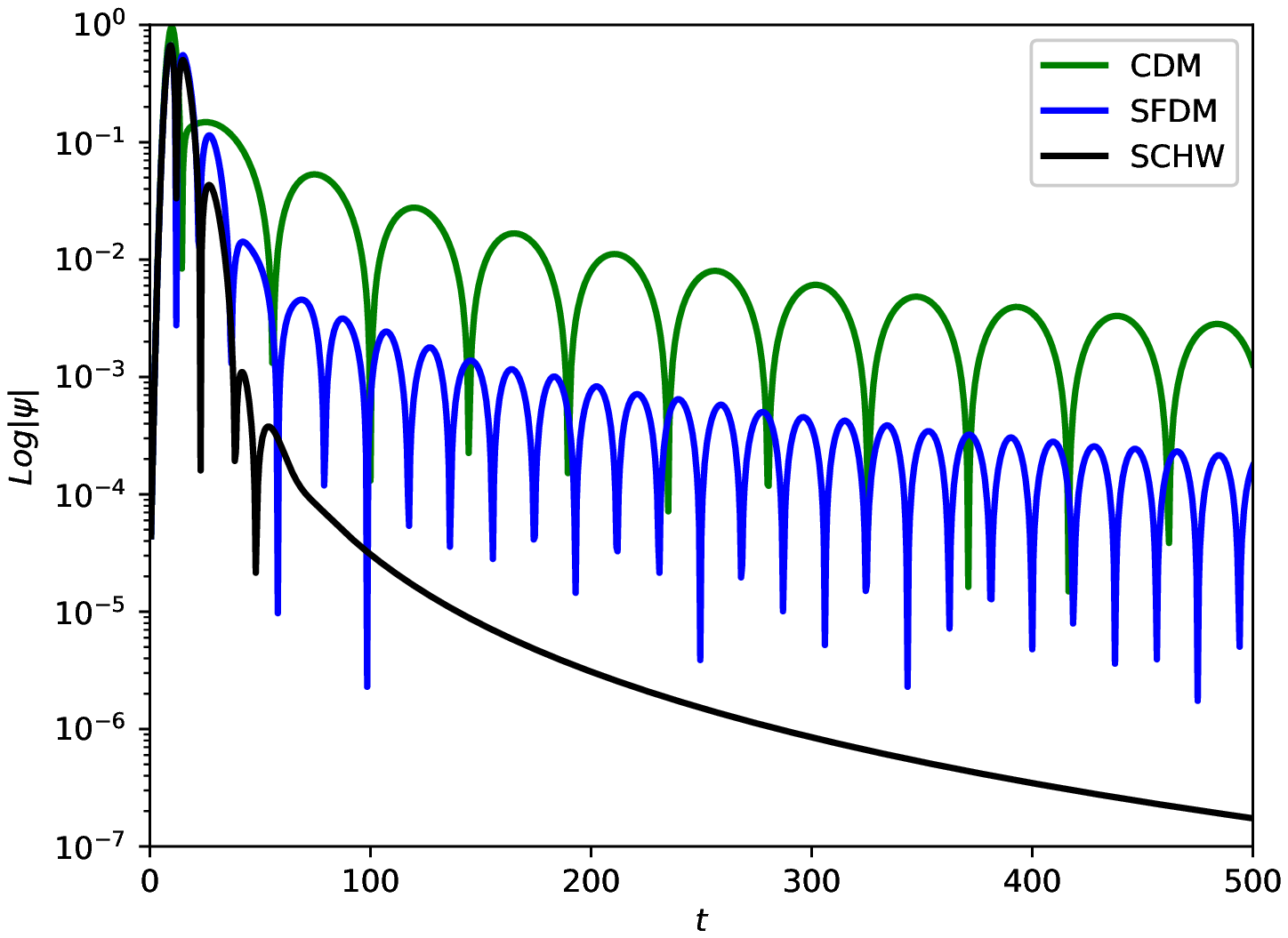}     
} 
\subfigure[$l=1$]{                                                
\label{fig:b}     
\includegraphics[width=0.6\columnwidth]{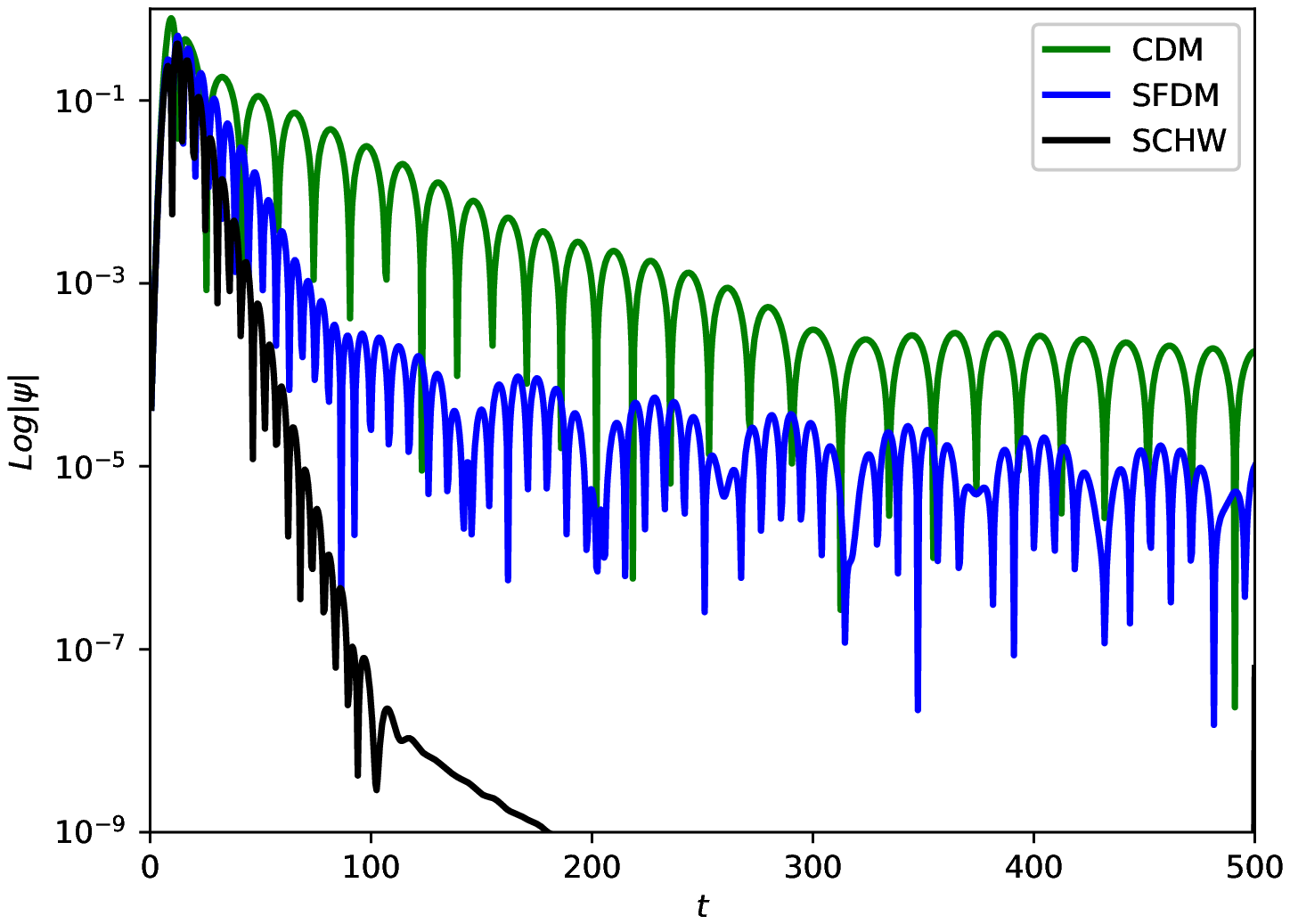}     
}    
\subfigure[$l=2$]{ 
\label{fig:b}     
\includegraphics[width=0.6\columnwidth]{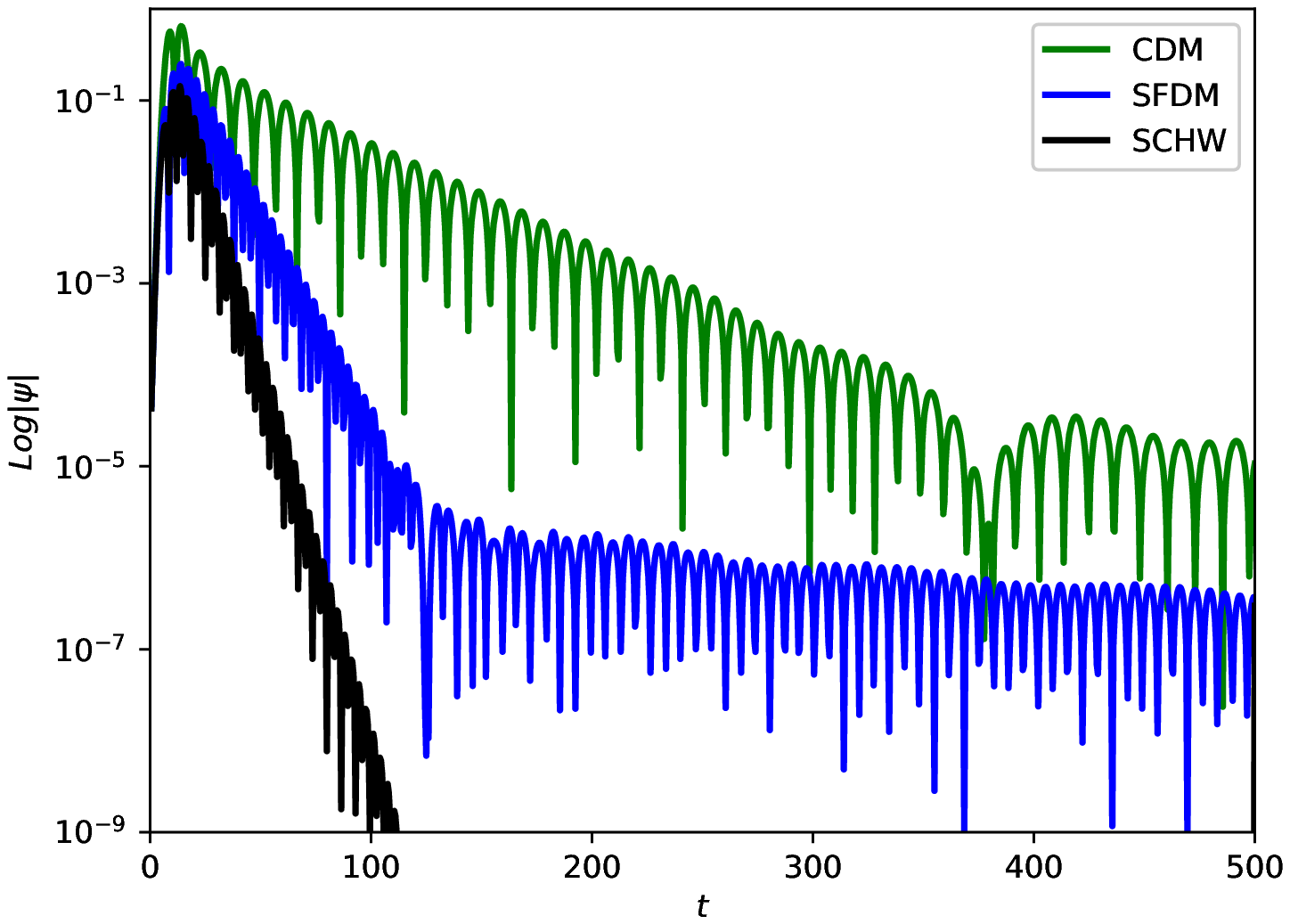}     
}   
\caption{Comparisons of quasinormal modes in the scalar fields with the different space-times.}     
\label{fig:13}   
\end{figure*}

\begin{figure*}
\centering  
\subfigure[$l=1$]{ 
\label{fig:b}     
\includegraphics[width=0.8\columnwidth]{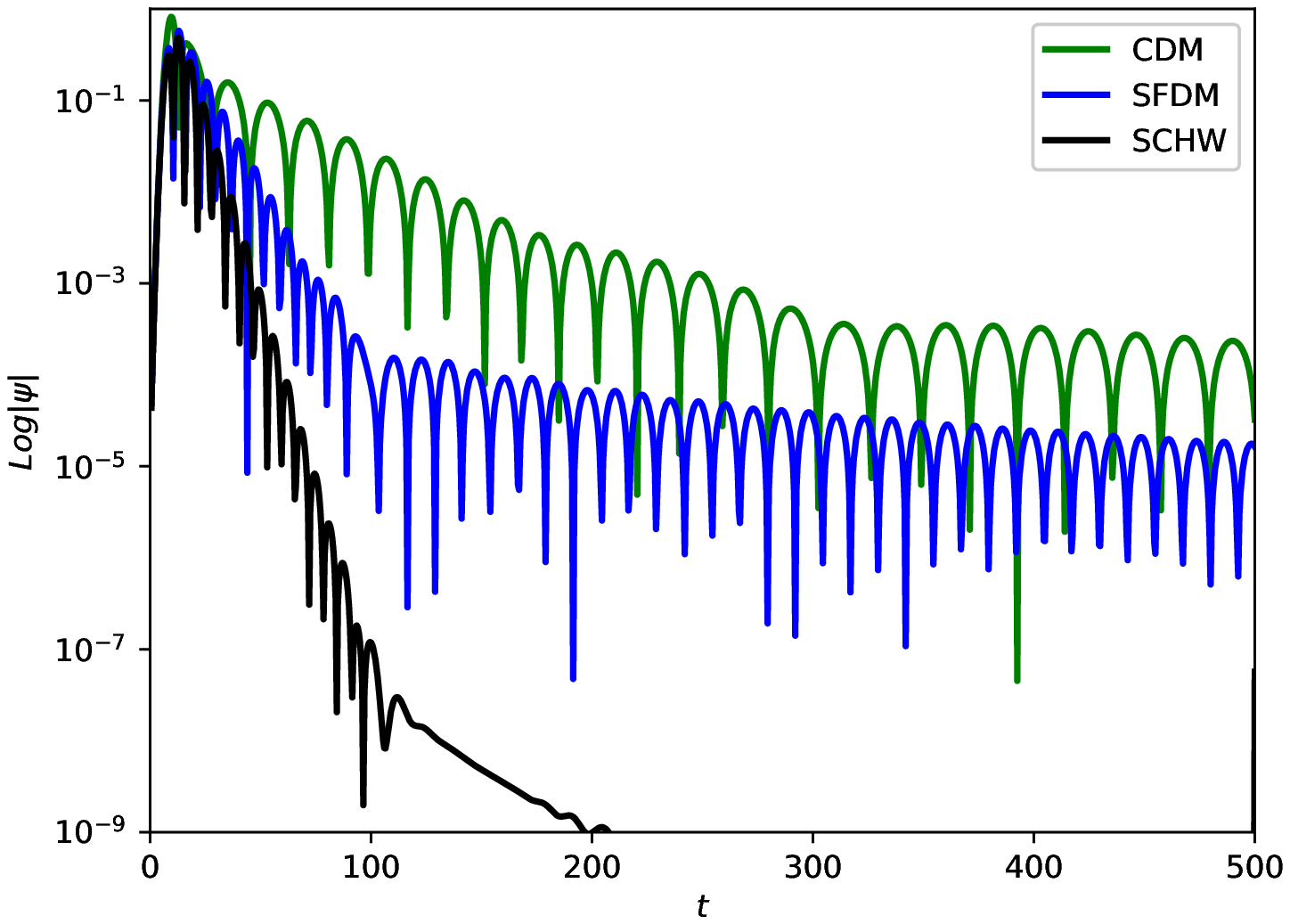}     
} 
\subfigure[$l=2$]{ 
\label{fig:b}     
\includegraphics[width=0.8\columnwidth]{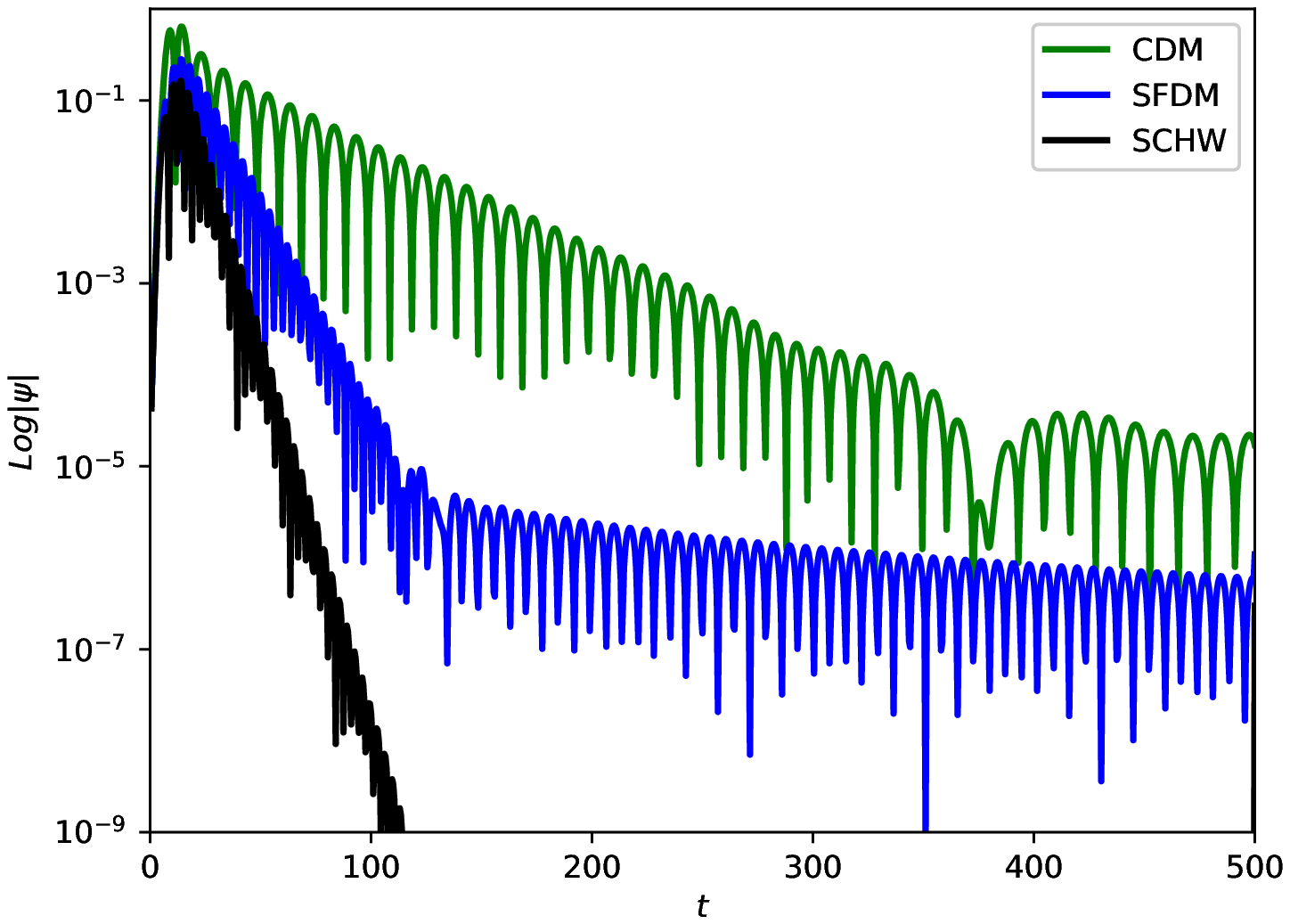}     
} 
\caption{Comparisons of quasinormal modes in the electromagnetic fields with the different space-times.}     
\label{fig:14}     
\end{figure*}

\begin{figure*}[t!]
\centering  
\subfigure[$l=2$]{ 
\label{fig:b}     
\includegraphics[width=0.8\columnwidth]{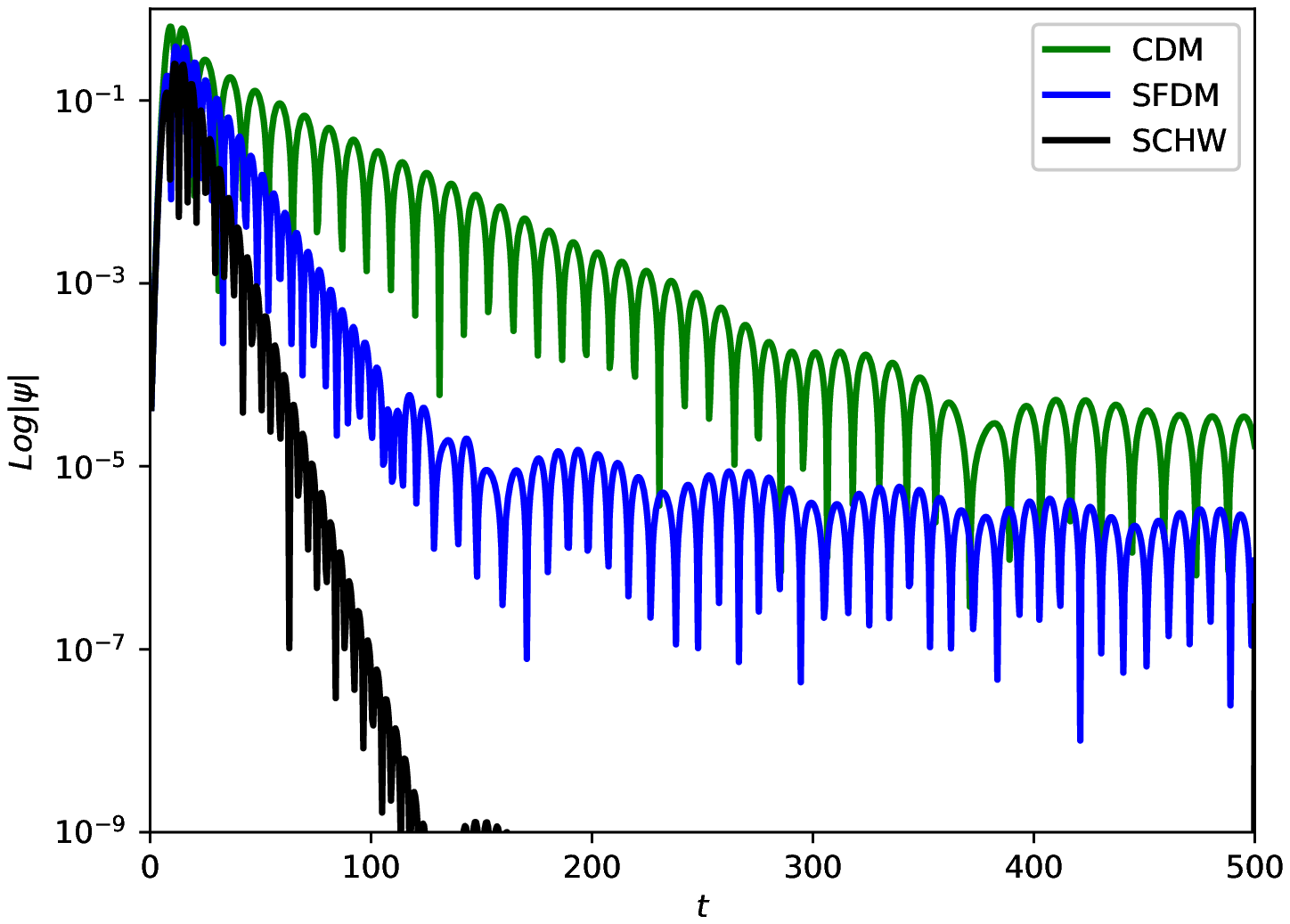}     
} 
\subfigure[$l=3$]{ 
\label{fig:b}     
\includegraphics[width=0.8\columnwidth]{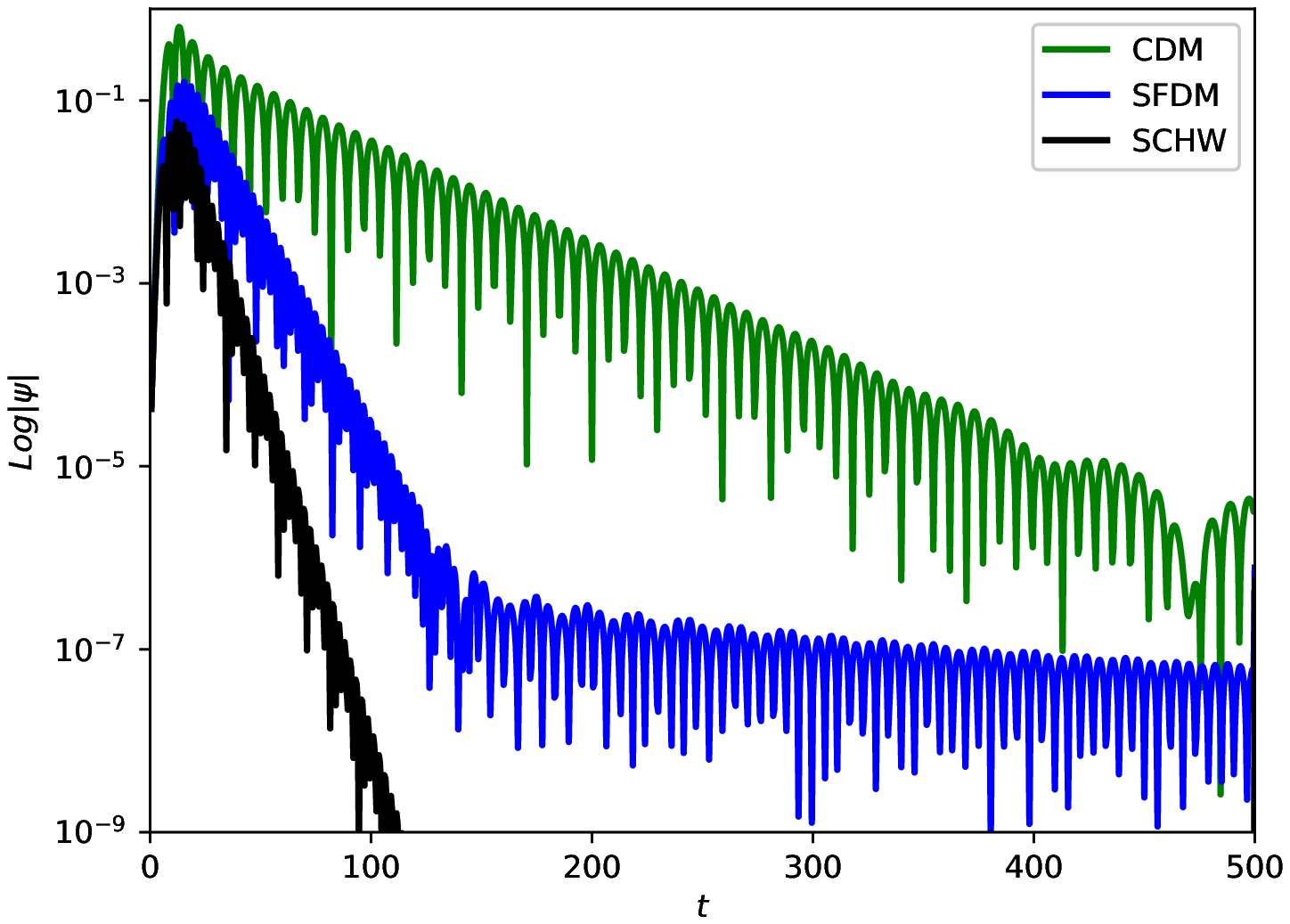}     
} 
\caption{Comparisons of quasinormal modes in the gravitational perturbation with the different space-times.}     
\label{fig:15}     
\end{figure*}

\begin{figure*}[t!]
\centering  
\subfigure[CDM]{ 
\label{fig:b}     
\includegraphics[width=0.6\columnwidth]{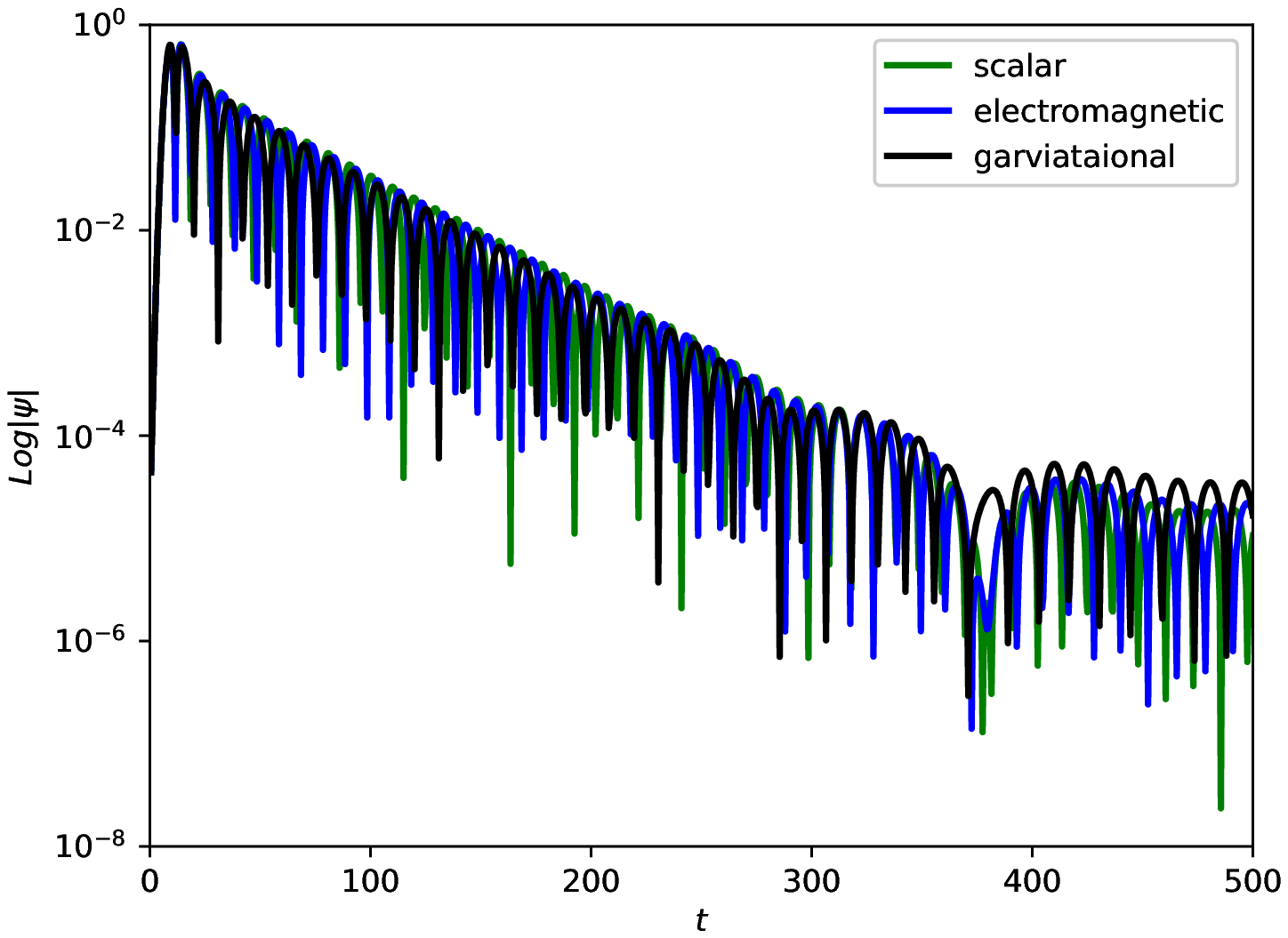}     
} 
\subfigure[SFDM]{ 
\label{fig:b}     
\includegraphics[width=0.6\columnwidth]{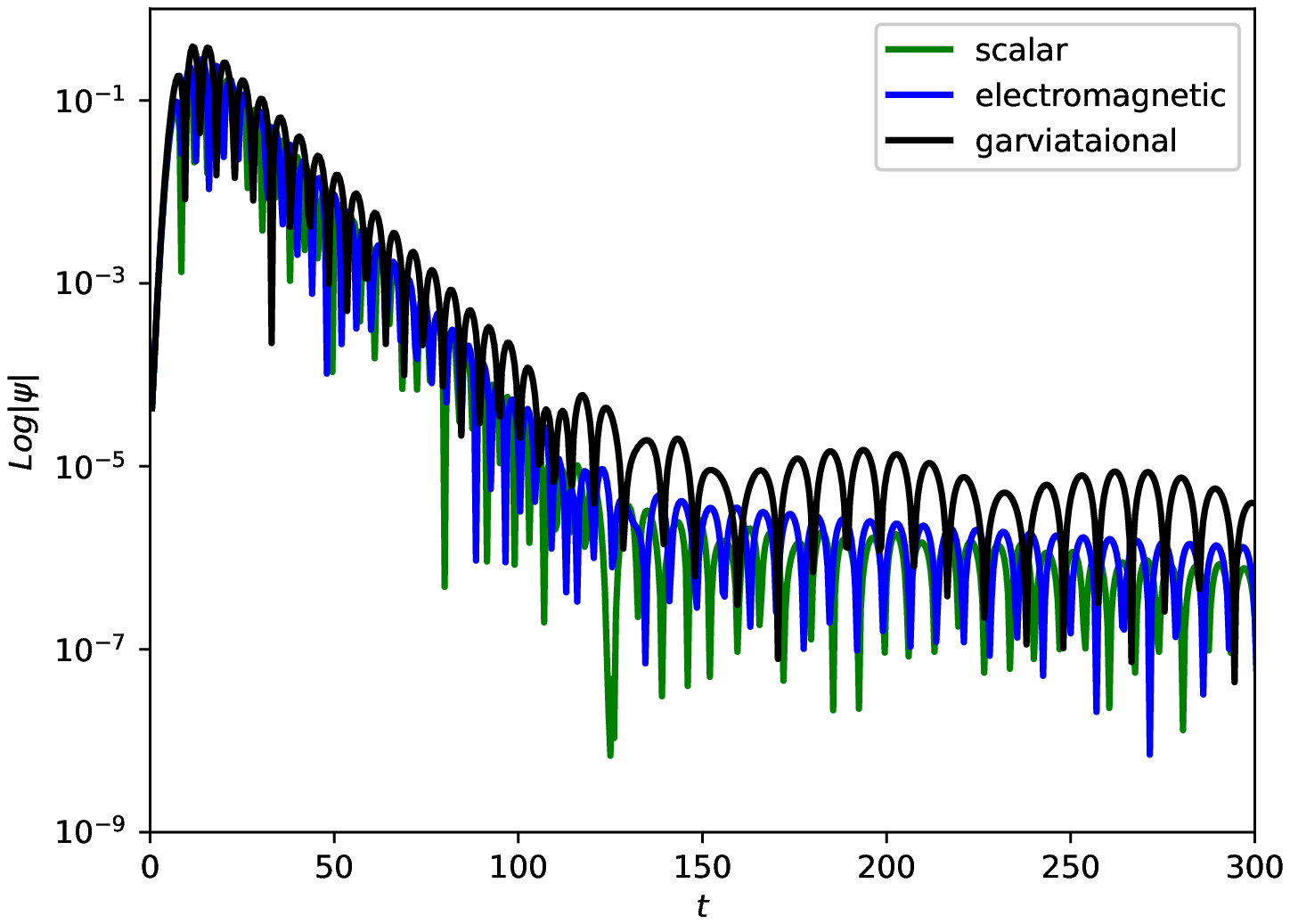}     
} 
\subfigure[SCHW]{ 
\label{fig:b}     
\includegraphics[width=0.6\columnwidth]{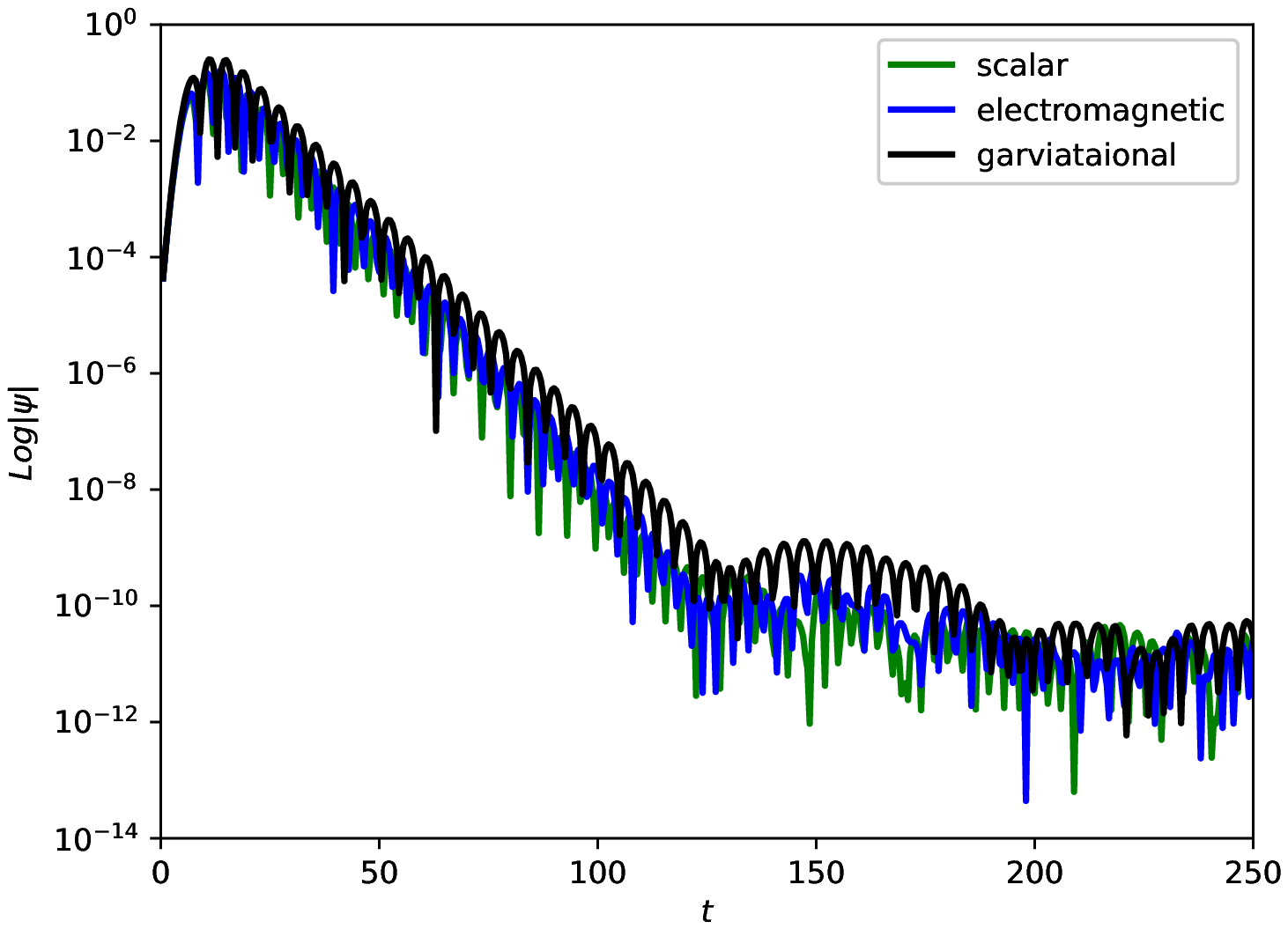}     
} 
\caption{Comparisons of quasinormal modes with the different backgrounds($l=2$).}     
\label{fig:16}     
\end{figure*}

From the data in Tables \ref{tab:1}-\ref{tab:3}, the results we obtained by the WKB method and the Prony method are in good agreement. These data indicate that the results we obtained are reliable. The reason for the error is that we consider that when using the Prony method to fit the frequency, it needs to artificially set the values of $t$ for the initial and final state. This is an accidental error, and averaging multiple measurements may effectively reduce the error. From the frequencies in Tables \ref{tab:1}-\ref{tab:3}, due to the fact that the effective potentials are positive, the imaginary parts of the frequencies in all Tables are all negative values, which indicates that there is a stable black hole solution in a dark matter halo. In addition, we found that the frequencies of SCHW are greater than SFDM and CDM, that of SFDM are greater than CDM. The imaginary parts of the frequencies are related to the attenuation and time of the QNMs, which correspond to the dynamical evolutions of QNMs. The QNM frequencies for SFDM attenuated faster than that of CDM, whereas QNMs of CDM are easier to be detected (Figs.\ref{fig:13}-\ref{fig:15}).

\indent On the other hand, when we set $ \rho $ and $R$ as constant values, the dynamical evolutions of CDM, SFDM, and SCHW are all related to the parameter $l$. Our research results also show that in a dark matter halo, both CDM and SFDM have QNMs ringing (Figs.\ref{fig:7}-\ref{fig:9}). In the case of CDM (the first panel in Fig.\ref{fig:9}), $t$ from $300$ to $400$, QNMs show fluctuations different from ringing. It is different from the case of SFDM.

For Schwarzschild black hole (Fig.\ref{fig:7}), QNMs appear after the initial phase. The QNMs ringing of CDM are between $10^{-1}$ and $10^{-3}$, that of SCHW are between $10^{-1}$ and $10^{-7}$. Figures \ref{fig:10}-\ref{fig:12} show that QNMs ringing time increases with increasing angular quantum number $l$. Tables \ref{tab:1}-\ref{tab:3} show that the imaginary part of QNM frequencies, which is related to the attenuation of QNMs, also increases with increasing angular quantum number $l$. With the increasing the angular quantum number $l$, the corresponding effective potentials will increase but QNMs will gradually attenuate. Figures \ref{fig:13}-\ref{fig:15} show that the comparisons of the three QNMs signal in the same condition. On the whole, QNMs of CDM are stronger than that of SFDM in the same condition, which is easier to be detected. In addition, black hole QNMs of CDM and SFDM in a dark matter halo are different from the Schwarzschild black hole, unlike Schwarzschild black hole with prominent power-law tail. Different kinds of models of dark matter can be distinguished by theirs QNMs.

Comparing these three different space-time backgrounds, QNMs ringing under the gravitational perturbation are greater than that of scalar field and electromagnetic field perturbations(Fig.\ref{fig:16}). This means that the gravitational radiation excited by gravitational perturbation is much larger than that excited by the external field. In the ways of black hole perturbations, gravitational perturbation may be more effective.

\section{Conclusions and Discussions}\label{sec:4}
As the characteristic sound of black holes, QNMs are one of the important means to understand black holes currently. QNM frequencies are not only related to the hairs of the black hole (mass, charge and angular momentum) but also may identify the existence of the black hole. So, in this paper, we study the black hole QNMs in a dark matter halo, and make comparisons with a Schwarzschild black hole. First, we consider the motion equation in a dark matter halo for scalar field, electromagnetic field and gravitational perturbation, and obtain corresponding effective potentials. Then we use the WKB method and the finite difference method for numerical work, and QNM frequencies and the dynamical evolutions of QNMs are obtained. Finally, based on the experimental data we obtained, the relevant research results are as follows:

(1) When the transcendental items are equal to constant $1$ in a dark matter halo, they will become a Schwarzschild black hole. It seems that in a certain situation, the black hole in a dark matter halo can degenerate to a Schwarzschild black hole.\\

(2) It can be seen from the distributions of the effective potential that the effective potentials of the dark matter halo increase with the increasing $l$, and the maximum values of the effective potentials are less than a Schwarzschild black hole. In a dark matter halo, when $r_*$ tends to negative infinity, their effective potentials are a positive value. When the $r_*$ tends to positive infinity, the effective potentials tend to $0$. By that moment, the black hole will no longer be affected by the dark matter halo and return to a stable state. \\

(3) QNMs are the inherent oscillation of a black hole under certain conditions, and its solution can be expressed as a purely outgoing wave at infinity and a purely ingoing wave at the event horizon, which becomes weaker with time, and finally returns to a balanced state. From the dynamical evolutions, the black hole QNMs of CDM and SFDM in a dark matter halo are different from a Schwarzschild black hole, unlike a Schwarzschild black hole with prominent power-law tail. So the different kinds of models of dark matter can be distinguished by their QNMs. QNMs of CDM show fluctuations different from ringing. It is different from the case of SFDM.\\

(4) The QNM signals of the dark matter halo occur after the initial phase. Then QNMs ringing dominates rapidly, and attenuates with the increasing parameter $l$. Due to the fact that the parameter $l$ is related to effective potentials, and QNMs are affected by the effective potentials. The effective potentials increase, whereas the QNMs become weaker. As parameter $l$ increases, the time of QNMs ringing becomes longer.\\

(5) The QNMs ringing of CDM is approximately between $10^{-1}$ and $10^{-3}$, that of SFDM is between $10^{-1}$ and $10^{-4}$. The Schwarzschild black hole is approximately between $10^{-1}$ and $10^{-7}$. The overall QNMs of CDM are greater than that of SFDM in the same condition, which is easier to be detected (Figs.\ref{fig:13}-\ref{fig:15}). This is consistent with the results we obtained by the frequencies in Tables \ref{tab:1}-\ref{tab:3}.\\

(6) Comparing these three different space-time backgrounds, QNMs ringing under the gravitational perturbation are greater than that of scalar field and electromagnetic field perturbations. This means that the gravitational radiation excited by gravitational perturbation is much larger than that excited by the external field. In the ways of black hole perturbations, gravitational perturbation also may be effective.\\

(7) From these conclusions, the distributions of different dark matter have different effects on black hole QNMs. In future studies, it may be possible to distinguish dark matter models by their special QNMs.\\

(8) Besides, QNM frequencies we obtained are in good agreement after fitting QNM data with the sixth-order WKB method and the Prony method.\\

In this paper, we study the case of dark matter halo. In fact, there may be a spike phenomenon in dark matter near the black hole\cite{P. Gondolo,L. Sadeghian,B.D. Fields}. In the case of dark matter spikes, its density will greatly increase, and its situation may be more complicated. Next, we will consider the case of dark matter spikes, and the studies based on QNMs may be checked in future gravitational wave plans.

\begin{acknowledgments}
We are very grateful to R. Moderski and M. Rogatko; R. A. Konoplya and A. Zhidenko; F. L. Carneiro and J. W. Maluf; C. Gundlach and R. H. Price; E. H. Djermoune; M. R. Osborne and G. K. Smyth for kindly providing us with useful code. We would also like to thank V. Cardoso; K. A. Bronnikov; E. George; D. R. Poulami; A. Chowdhury for helpful correspondence. This research was funded by the National Natural Science Foundation of China (Grant No.11465006 and No.11565009) and the Natural Science Special Research Foundation of Guizhou University (Grant No.X2020068).
\end{acknowledgments}

\nocite{*}

\end{document}